\documentclass[12pts,fleqn,usenatbib,useAMS]{mnras}

\usepackage{amsmath}
\usepackage{color}
\usepackage{subfigure}
\usepackage{float}

\RequirePackage{fixltx2e}

\def\la{\mathrel{\hbox{\rlap{\hbox{\lower4pt\hbox{$\sim$}}}\hbox{$<$}}}}

\def\ga{\mathrel{\hbox{\rlap{\hbox{\lower4pt\hbox{$\sim$}}}\hbox{$>$}}}}

\usepackage{journal_names}
\usepackage{color}

\usepackage{url}

\newcommand\farcss{\mbox{$.\!\!\!^{\prime\prime}$}}
\def\farcm{\mbox{.\kern -0.5ex\raisebox{.6ex}{\scriptsize$\prime$}}}

\def\farcss{
 \mbox{ 
  \kern  0.13ex. 
   \kern -0.95ex\raisebox{.6ex}{\scriptsize$\prime\prime$}
  \kern -0.1ex
 }
}

\usepackage{graphicx,times}

\usepackage{float}

\title[]{Planet population synthesis driven by pebble accretion in cluster environments}
\author[~Ndugu et al.]
{\parbox{\textwidth}{N.~Ndugu$^{1}$\thanks{E-mail: ndugu.nelson89@gmail.com},
B.~Bitsch$^{2}$\thanks{E-mail: bert@astro.lu.se},~ 
E.~Jurua$^{1}$\thanks{E-mail: ejurua@must.ac.ug}~} \vspace{0.4cm}\\
\parbox{\textwidth}{$^{1}$Department of Physics, Mbarara University of Science and Technology, Mbarara, Uganda\\
$^{2}$Lund Observatory, Department of Astronomy and Theoretical Physics, Lund University, 22100 Lund, Sweden} }

\begin{document}

\pagerange{\pageref{firstpage}--\pageref{lastpage}} \pubyear{2017}

\maketitle

\label{firstpage}

\begin{abstract}
The evolution of protoplanetary discs embedded in stellar clusters depends on the age and the stellar density in which they are embedded. Stellar clusters of young age and high stellar surface density destroy protoplanetary discs by external photoevaporation and stellar encounters. Here we consider the effect of background heating from newly formed stellar clusters on the structure of protoplanetary discs and how it affects the formation of planets in these discs. Our planet formation model is build on the core accretion scenario, where we take the reduction of the core growth time-scale due to pebble accretion into account. We synthesize planet populations that we compare to observations obtained by radial velocity measurements. The giant planets in our simulations migrate over large distances due to the fast type-II migration regime induced by a high disc viscosity ($\alpha=5.4 \times 10^{-3}$). Cold Jupiters ($r_{\rm p}>1$ AU) originate preferably from the outer disc, due to the large scale planetary migration, while hot Jupiters ($r_{\rm p}<0.1$ AU) preferably form in the inner disc. We find that the formation of gas giants via pebble accretion is in agreement with the metallicity correlation, meaning that more gas giants are formed at larger metallicity. However, our synthetic population of isolated stars host a significant amount of giant planets even at low metallicity, in contradiction to observations where giant planets are preferably found around high metallicity stars, indicating that pebble accretion is very efficient in the standard pebble accretion framework. On the other hand, discs around stars embedded in cluster environments hardly form any giant planets at low metallicity in agreement with observations, where these changes originate from the increased temperature in the outer parts of the disc, which prolongs the core accretion time-scale of the planet. We therefore conclude that the outer disc structure and the planet's formation location determines the giant planet occurrence rate and the formation efficiency of cold and hot Jupiters.
\end{abstract}


\section{Introduction}
Most stars are born in stellar clusters \citep{Lada2003} and spectroscopic observations have revealed that young stars are rich in circumstellar discs \citep{Lada2000}. The longevity of the protoplanetary disc is a strong function of its environment  \citep{Pfalzner2015,Pfalzner2016}. Discs in stellar clusters face two main hurdles, photoevaporation and stellar encounters which limit both the lifetimes and sizes of discs. Photoevaporation sweeps out materials from the discs rendering shorter lifetimes where the effect is stronger for young stellar clusters \citep{Johnstone1998}.

Discs in massive stellar clusters encounter star-disc and disc-disc encounters, which destroy the disc by accreting it onto the central star or capturing part of it by the perturbing star \citep{Susan2012}, making planet formation in young and massive stellar clusters extremely difficult. Nevertheless, \cite{Malavolta2016} observed a multi planet system in the open stellar cluster M44. Stellar clusters can have hot background temperatures of 40 - 100 K \citep{Krumholz2006}, which could last during the entire lifetimes of protoplanetary discs until OB stars go supernova. For example, in NGC1333 the highest heating influence of B stars is 40 K \citep{Hatchell2013}. \cite{Spaans1995} also additionally found that young stellar objects can have a heating influence of few hundred Kelvins up to even 1000 AU. Embedded discs are thus different from discs of single field stars. The background heating of a protoplanetary discs by a neighbouring star population is attributed to both accretional heating and intrinsic heating of the star population. Here, accretional heating trumpets at the early phase of the star and disc formation process, while intrinsic heating becomes very important within the vicinity of massive stars and at later times.

Stars in a loosely bound stellar clusters of stellar density less than $\sim  2 - 3 \times 10^{3} \rm pc^{-2}$ have only a few encounters, like in NGC6791 \citep{Rosotti2014}. This implies that stars in less massive stellar clusters are to a close approximation in an isolated star field but feel the ambient stellar cluster heating. The mentioned stellar density specification for a stellar encounter is well within the definition of stellar cluster in \cite{Krumholz2006}. The goal of our work is thus to investigate the influence of a stellar cluster environment on planet formation by focusing on the influence of the stellar cluster on the thermal properties of the disc. We therefore do not model any gravitational encounters between different stars or external photoevaporation, which can significantly alter the disc properties, but rather consider the background temperatures of stellar clusters.

We expand the solar like disc evolution model of \cite{Bitsch} to mimic the disc evolution model for stellar clusters by setting different background temperatures. In the disc, we implant planetary seeds that have reached the pebble transition mass, indicating that the seeds can accrete in the efficient Hill accretion regime \citep{Lambrechts2014}. The planet accretes pebbles rapidly and when it reaches the pebble isolation mass, at which the pebble flux is shut-off due to interactions of the planet with the disc blocking the flux of pebbles, after several 100 kyrs \citep{Lamb2014}, it contracts an gaseous envelope \citep{Pisso} and then undergoes run-away gas accretion, when the planetary envelope becomes more massive than the planetary core, where we follow the accretion rates derived by \cite{Machida}.

Planets migrate through the disc during their entire growth. Small planets that do not perturb the disc significantly, migrate in type-I fashion \citep{Paardekooper2011}, while massive planets open gaps in the disc and migrate in type-II fashion \citep{Lin}. Our planet growth and migration model is based on the work of \cite{Bitsch2015}, which we expand in order to synthesize planet populations by probing different initial conditions (e.g. metallicity, initial starting semi-major axis of the planetary seeds). The resulting planet populations are then compared with observations to test our predictions, where we focus especially on the metallicity correlation for the formation of giant planets, namely that giant planets form preferably around stars with large metallicity \citep{Fischer,J2010}. The metallicity correlation was also studied in planet population synthesis simulations, where the planetary core was growing through the accretion of planetesimals \citep{Ida2004, Matsuo2007, Mordasini2009}, where the synthetic planet population shows a less steep trend compared to observations.

The work is structured as follows. In section 2, the different methods used for disc evolution, pebble and gas accretion, and for planet migration and planet population synthesis are described. The disc structure, growth tracks, and planet population synthesis simulation and their comparison to observations are presented in section 3. Then finally, the work is summarized in section 4. 

\section{Methods}

\subsection{Background heating of stellar clusters}

We note here that we do not include the effects of external photoevaporation onto the disc structure, even though external photoevaporation from close by O-stars can significantly change the disc structure and disc lifetime (e.g. \cite{Henney02}), but even within low mass star forming regions, external photoevaporation can play a role \citep{Haworth17}. Here we investigate instead the change of the temperature field from the background heating of stellar clusters, which originates from either accretion luminosity of young stars or from intrinsic stellar luminosity.

During the onset of star formation, the accretion rate onto the new stars is very high, implying very high accretion luminosities, which heat the dust around the central star to very high temperatures. However, as the star ages, the accretion rate reduces by orders of magnitude \citep{Hartmann2016}, resulting in reduced accretion rates and accretion luminosities.

Dust around central stars in stellar populations is also heated due to the intrinsic luminosity of massive stars. This heating dominates at late stages of the disc evolution, when accretion is very low. At this stage, heating of a nearby protoplanetary disc is due to the intrinsic luminosity and not through the accretional luminosity. The background heating in a stellar cluster is due to the interplay of accretional and intrinsic luminosity of stars. Accretional luminosity is dependent on the accretion rate while intrinsic luminosity scales with the mass of the stars.

Star populations in a cluster are most often assumed to have the same age. In reality this is not true because within a stellar cluster, age spreads of stars occur, for example LH 95 hosts an age spread of 2.8-4.4 Myr \citep{DaRio2010}. For simplicity of the background heating estimate, we did not consider the age spread of star population in a cluster but assumed a uniform age of star population. Therefore in our calculation of the ambient stellar cluster heating onto a central star, the adjective 'estimate' is the key.

\begin{figure}
 \centering
 \includegraphics[width=3.5in]{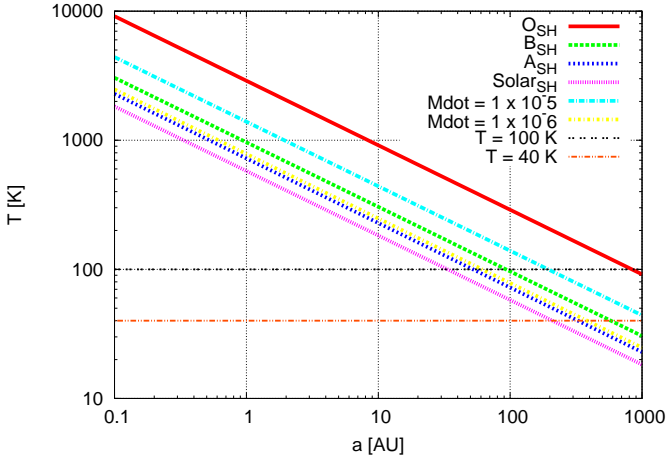}
 \caption{Background heating variation with orbital distance to various heating sources. Protoplanetary discs close to massive stars experience very high background heating. The orange and the black lines display our assumptions of stellar cluster background heating of 40 K and 100 K and for which sources this can be maintained. Here the index SH stands for Stellar heating and $\dot{M}$ represents the accretion rates in Solar mass per year of accreting discs. O, B and A are the spectral classes of stars considered.}
 \label{fig:0}
\end{figure}

For a central host star embedded in a population of solar like stars, the contribution of background heating onto the host star is shown in Fig.~\ref{fig:0}. Here we make the simplifying assumption that the heating and cooling of the dust grains happens at the same wavelength, whereas in reality stellar irradiation peaks in the optically, while dust grains emit in the infrared (e.g. \cite{Chiang1997}). To obtain Fig.~\ref{fig:0}, the accretional luminosity is calculated as
\begin{equation}
  \label{eq:Lacc}
  L_{\rm acc} = \frac{GM_{\star}\dot{M}_{\star}}{R_{\star}} \ .
 \end{equation} 
 Here $G$ marks the gravitational constant, $R_\star$ the stellar radius, $M_\star$ the stellar mass and $\dot{M}_\star$ the accretion rate onto the central star. The intrinsic stellar luminosities were set to the luminosities of O, B, A and solar like stars.  The corresponding luminosities were then translated into  a temperature - distance relation displayed in Figure~\ref{fig:0} using the expression
\begin{equation}\rm
  \label{eq:Tacc}
  T = \left(\frac{L}{4{\pi}\sigma R^2}\right)^{\frac{1}{4}} \ .
 \end{equation}
L stands for luminosity (intrinsic or accretional luminosity), $\sigma$ is the Stefan-Boltzmann constant and R is the distance from the star. In Fig.~\ref{fig:0}, we over-plot both the stellar irradiation heating and the accretional heating contribution. It can been seen from Fig.~\ref{fig:0} that the influence of accretional heating decreases as the star ages (lower accretional rates) and close massive stars heat the discs to higher temperature. For discs distantly located from the heating stars (1000 AU), the background heating of 100 K is not easily achievable from the estimates in Fig.~\ref{fig:0}.

In stellar clusters with low densities of $\sim  2 - 3 \times 10^{3} \rm pc^{-2}$, stars have only few encounters \citep{Rosotti2014}, but the stellar density in these low density clusters, would be a factor of a few too low to justify background temperatures of 40 K using the temperature estimates of Fig.~\ref{fig:0}\footnote{{\bf We have converted the stellar density to a mean star-star distance, giving us estimates about the background temperature using Fig.~\ref{fig:0}.}} Nevertheless, \cite{Spaans1995} stressed that heating of a few hundred Kelvins are attainable at 1000 AU for stars surrounded by young stellar objects of radiation field of 10000 K.  Background heating of few hundred K were also noticed in \cite{Krumholz2006}. From Fig.~\ref{fig:0}, the background heating of 100 K seems however only to be possible for a location at 100-1000 AU from the massive host stars. Therefore, we state here from our rough estimate that the higher background heating of 100 K is only attainable if massive stars are in the cluster, where then external photoevaporation plays a role \citep{Henney02}. While a background heating of 40 K can even be obtained by intrinsic heating of close massive stars and by the accretion luminosity of young stars. Trailing our background heating calculation, 100 K is an extreme case, implying that the more realistic  estimate is 40 K. In the following sections we apply such background heating of stellar clusters to our disc structure, simulate planet populations using the developed disc structure and test the importance of such discs in mimicking the observed gas giants populations.

\subsection{Disc Structure and Evolution}

Protoplanetary discs comprise of dust and gas. Dust particles inside the disc stick and coagulate to form larger pebbles which eventually form planetesimals, for example through the streaming instability (for review see \cite{Johansen2014}). Planetesimals eventually accrete more planetesimals or pebbles to form the even larger planetary cores, which gradually accrete gas when they becomes massive enough. The growth via pebbles and planetesimals depends on the local disc structure, where especially the aspect ratio $H/r$ and the radial pressure gradient play important roles, highlighting the importance of the protoplanetary disc structure for forming planets \citep{Bitsch2015,Bitsch2016}.

In this study, our disc structure is not a simple power law as in the MMSN (Minimum Mass Solar Nebulae), but is determined by an equilibrium between stellar and viscous heating with radiative cooling. The resulting equilibrium disc structure is determined by the opacity profile of the disc as this determines the cooling rates. Here our underlying disc model follows \cite{Bitsch}.

The disc dust mass composition is made of the following ingredients; $\rm CO$ (55.9 $\%$), $\rm CO_{2}$ (9.8 $\%$), $\rm H_{2}O$ (14.3 $\%$) and Silicate (20.0 $\%$) with evaporation temperatures of 20 K, 70 K, 170 K, 1500 K, respectively, but the disc model only consists of $\rm H_{2}O$ and Silicate because these are the two species believed to contribute to opacity transitions, in contrast to $\rm CO$ and $\rm CO_{2}$, where no transition at the respective ice lines is observed \citep{Qi2013}. The disc's relative abundances follow the findings of \cite{Asplund2009} for solar composition. At the ice line of each ingredient, the fraction of the respective dust ingredient is evaporated, reducing the pebble flux. This implies several ice lines in the disc, where the disc model will take the transitions at the Silicate line and $\rm H_{2}O$ line into account      producing bumps in the disc structure \citep{Bitsch}. The disc structures are not changed at the $\rm CO$ and $\rm CO_{2}$ ice lines, as those are only minimal transition 
of opacity. We used the disc time evolution presented in \cite{Hartmann}, which is summarized as 
 \begin{equation}
  \label{eq:lifetime}
  \log_{10}\left(\frac{\dot{M}_\star}{\rm M_{\odot}/yr}\right) = -8.00 - 1.40\log_{10}\left(\frac{t + 10^{5} \rm yr}{10^{6}\rm yr}\right).
 \end{equation} 
Here, $M_{\odot}$ is the solar mass. The stellar accretion rate $\dot{M}_\star$ is related to the viscosity $\nu$ and the gas surface density $\Sigma_{\rm G}$ as
\begin{equation}
 \label{eq:accretionrate}
  \dot{M}_\star = 3{\pi}{\nu}{\Sigma_{\rm G}} \ .
\end{equation}
We use here the $\alpha$ viscosity with $\alpha = 0.0054$ as in \cite{Bitsch}.  We then extend the disc structure calculated in \cite{Bitsch} to mimic the stellar cluster environment. This modification holds for stellar clusters that have few encounters (for example NGC6791 \citep{Rosotti2014}). Discs in dense stellar clusters experience rampant stellar encounters, which consequently destroys the discs. However, in our model we did not consider stellar encounters in the cluster, but rather considered a background temperature of 40 or 100 K. The assumption in \cite{Krumholz2006} showed that stellar clusters born from molecular cloud can have background temperature in the range of 40 k - 100 K. The background temperatures hold as long as the stellar cluster is younger than 10 Myr when supernovae clear the remaining gas of the clusters, which interestingly is within the lifetime of protoplanetary discs \citep{Mamajek2009}. Older stellar clusters have lower background temperatures due to fewer remaining stars. The disc model feature $ c_{\rm s} < v_{\rm esc}$, meaning that the disc is bound to the central star, even for large cluster background temperatures.

\subsection{Disc profile}

\begin{figure}
 \centering
 \includegraphics[width=3.5in]{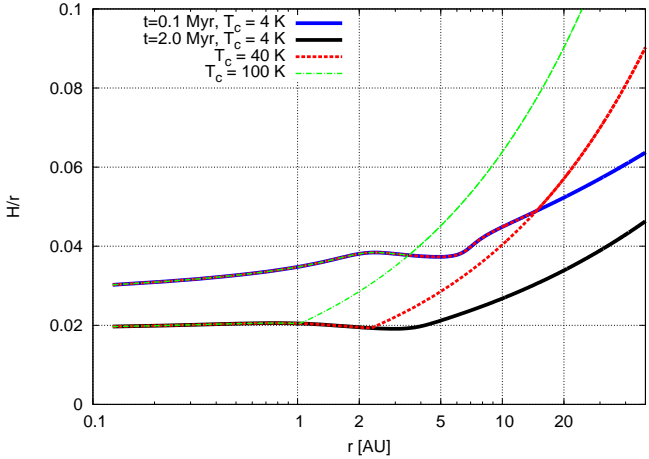}
 \caption{Aspect ratio profile for discs with age $t =  0.1$ Myr (blue) and  $t = 2$ Myr (black) around an isolated host star and in stellar clusters with cluster temperatures of 40 K (red) and 100 K (green) for the two ages. In the very inner regions of the disc, viscous heating dominates also over the cluster heating, which is why the disc profiles follow the same trend independent of the background cluster temperature. This is independent of the disc age, but as the disc cools in time, the viscously dominated regions move inwards. For cluster temperatures of 40 K (red) the transition between the viscously dominated part and the cluster heating dominated part of the disc is farther away from the star as for the cluster temperature of 100 K (green).}
 \label{fig:1}
\end{figure}

Fig.~\ref{fig:1} shows the aspect ratio profile for discs in an isolated star field and a clustered star field. The pebble isolation mass (Equation \ref{eq:Iso}) and the pebble surface density (Equation \ref{eq:pebb}) are directly related to the aspect ratio of protoplanetary discs. Additionally, the migration rates depend on the aspect ratio of the disc, making the aspect ratio profile a strong constraint on planet formation models. From Fig.~\ref{fig:1}, it can be seen that clustered fields set a high aspect ratio at the outskirts of the disc, compared to isolated field stars. This is due to the stellar cluster dominance over stellar heating of the central star making the outer rim of the disc hotter than for the isolated stars. The hotter outer disc then translates into higher aspect ratios. Additionally, the stellar cluster dominance is pushed to the disc's inner part when the cluster is hotter. The stellar cluster dominance also scales with the accretion rate of the disc. Accretion rates directly relate with the disc's age; young discs have high accretion rates while old discs have low accretion rates. Discs with higher accretion rates are hotter than discs with lower accretion rates due to viscous heating and therefore the influence of stellar clusters on the disc structure increases with time.

\subsection{Planet formation and migration model}

We grow planets in the core accretion fashion, where pebbles are accreted onto a core, where eventually runaway gas accretion is triggered once the pebble isolation mass is reached and the gaseous envelope contracts. Pebbles with the Stokes number 
\begin{equation}
  \label{eq:Stocknumber}
  \tau_{\rm f} = \frac{{\rho_{\rm {\bullet}}} R}{ { \rho_{\rm g}} H_{\rm g}},
 \end{equation} are accreted onto cores that have reached 
 the pebble transition mass \citep{Lambrechts2012, Lambrechts2014} 
\begin{equation}
\label{eq:transitionmass}
 M_{\rm t} = \sqrt{\frac{1}{3}}\frac{\left({\eta}{\upsilon_{\rm k}}\right)}{G \Omega_{\rm k}}.
\end{equation}
Here, $ \rho_{\rm{\bullet}}$ is the material density,  $R$ is the particle radius, $\rho_{\rm g}$ is the gas density, $ H_{\rm g}$ is the gas scale height, ${\eta}$ is the pressure support, G is the gravitational constant, $ \upsilon_{\rm k}$ is the Keplerian speed and $\Omega_{\rm k}$ the Keplerian frequency.
Initially, the core accretes pebbles in a 3D fashion following the work of \cite{Lambrechts2012}, because the Hill radius is smaller than the pebble scale height for initially low mass cores. However, as the core attracts more pebbles, it becomes more massive and accretes in a 2D manner because $r_{\rm H}>H_{\rm peb}$ \citep{Morbidelli} with growth rate 
\begin{equation}
 \label{eq:2Daccretion} 
 \dot{M}_{\rm planet} = 2{\left(\tau_{\rm f}/0.1\right)^{2/3}}{r_{\rm H}} {v_{\rm H}}\Sigma_{\rm peb} \ .
\end{equation}
Here $r_{\rm H}$ denotes the planetary Hill radius, $v_{\rm H}$ the Hill speed at which the particles enter (given by $v_{\rm H} = r_{\rm H} \Omega_{\rm K}$) and $\tau_{\rm f}$ is the Stokes number of the pebbles. The pebble surface density is given by
\begin{equation}
 \Sigma_{\rm peb} = \sqrt{\frac{2\dot{M}_{\rm peb}\Sigma_{\rm g}}{\sqrt{3}\pi{\epsilon_{\rm p}}{r_{\rm p}}\upsilon_{\rm k}}} \ ,
 \label{eq:pebb}
\end{equation}
where $r_{\rm p}$ is the orbital distance of the planet. The pebble flux is given by
\begin{equation}
\label{eq:pebbleflux}
 \dot{M}_{\rm peb} = 2{\pi}r_{\rm g}\frac{dr_{\rm g}}{dt}\left(Z_{\rm peb}\Sigma_{\rm g}(r_{\rm g}\right) \ .
\end{equation}
Here $Z_{\rm peb}$ is the pebble metallicity, $\Sigma_{\rm g} (r_{\rm g})$ is the gas surface density at the location of the pebble production line $r_{\rm g}$ given by
\begin{equation}
 r_{\rm g} = \left(\frac{3}{16}\right)^{\frac{1}{3}}\left(GM_{\rm \star}\right)^{\frac{1}{3}}\left({\epsilon_{\rm D}}{Z_{\rm peb}}\right)^{\frac{2}{3}}t^{\frac{2}{3}}
\end{equation}
and 
\begin{equation}
 \frac{dr_{\rm g}}{dt} = \frac{2}{3}\left(\frac{3}{16}\right)^{\frac{1}{3}}\left(GM_{\rm \star}\right)^{\frac{1}{3}}\left({\epsilon_{\rm D}}{Z_{\rm peb}}\right)^{\frac{2}{3}}t^{-\frac{1}{3}} \ ,
\end{equation}
where $ M_{\rm \star}$ is the stellar mass, $\epsilon_{\rm p}$ and $ \epsilon_{\rm D}$ are the pebble and dust coagulation efficiency with values 0.5 and 0.05, respectively. It should be noted that at the ice lines, the pebble flux is significantly reduced by the factor of the corresponding volatile ingredient. Here we do not take the reduction of the pebble size into account as in \cite{Morbidelli}, where a reduced pebble size is one of the ingredients to explain the dichotomy between the terrestrial and giant planets of the solar system. This is the standard pebble accretion paradigm, where $\epsilon_{\rm p}$ and $ \epsilon_{\rm D}$ are values corresponding to detailed coagulation simulations (e.g \cite{Birnstiel2011}). Within the framework of \cite{Lamb2014}, the changes of accretion rate could only happen due to changes in coagulation efficiency, which is not motivated by the simulations of grain growth.

We note here that drift limited solution of pebble growth and drift produces pebble surface densities that are a factor of a few lower compared to observations. We therefore incorporate a pebble flux that corresponds to the pebble flux of a protoplanetary disc unperturbed by viscous disc decay outside of 50 AU, consistent with viscous disc evolution \citep{Baillie2015}. More details of this mechanism can be found in Bitsch et al. (2017, in prep.).

When the planet reaches a certain mass, it changes the gas pressure outside of its orbit due to a partial opening of a planetary gap, which halts pebble accretion onto the core and  gas accretion commences \citep{Lamb2014}. The pebble isolation mass is given by \citep{Lamb2014}
\begin{equation}
 M_{\rm iso} \approx 20\left(\frac{H/r}{0.05}\right) {\rm M}_{\rm E} \ .
 \label{eq:Iso}
\end{equation}
This stresses the importance of the disc structure for calculations of planetary growth.

At the pebble isolation mass, the gas envelope contracts over a long time while accreting some gas as long as $ M_{\rm env} < M_{\rm core}$ at a rate given by \citep{Pisso,Bitsch2015}
 \begin{eqnarray}
\label{eq:envelope}
 { \dot{M}_{\rm gas}} & = &{0.002}f^{-2}\left(\frac{ \kappa_{\rm env}}{1 \rm cm^{2}/g}\right)^{-1}\left(\frac{ \rho_{\rm C}}{5.5 \rm g/cm^{3}}\right)^{-\frac{1}{6}} \left(\frac{\rm M_{\rm c}}{\rm M_{\rm E}}\right)^{\frac{11}{3}}\nonumber \\ &&\left(\frac{M_{\rm env}}{0.1 \rm M_{\rm E}}\right)^{-1} \left(\frac{T}{81\rm K}\right)^{-0.5}\frac{\rm M_{\rm E}}{\rm Myr}
\end{eqnarray}
where $f$ is a normalization factor to map the accretion rate onto numerical and analytical results given by 0.2 \citep{Pisso} and $\kappa_{\rm env} = 0.05$ cm$^2$/g is the planet's envelope opacity very similar to the study by \cite{Movshovitz}. $\rho_{\rm C} $ is the core density and we assumed $\rho_{\rm C}$ = 5.5 $\rm{g/cm^{3}}$. The planet continues to contract its envelope until $M_{\rm c} = M_{\rm env}$. When $ M_{\rm c} < M_{\rm env}$, rapid gas accretion is triggered and the gas accretion rate is given by the minimum of \citep{Machida}
\begin{equation}
 \dot{M}_{\rm gas,\rm low} = 0.83\Omega_{\rm k}\Sigma_{\rm g}H^{2}\left(\frac{r_{\rm H}}{H}\right)^{\frac{9}{2}}
\end{equation}
and 
\begin{equation}
 \dot{M}_{\rm gas,\rm high} = 0.14 \Omega_{\rm k}\Sigma_{\rm g} H^{2}.
\end{equation}
Low mass planets migrate in the type-I fashion following the analytical torque formula in \cite{Paardekooper2011}, where the total torque, $ \Gamma_{\rm tot} $ acting on the planet is given by 
\begin{equation}
    \label{eq:tottorque}
   \Gamma_{\rm tot} = \Gamma_{\rm L} + \Gamma_{\rm C}.
\end{equation}
$ \Gamma_{\rm L}$ and $ \Gamma_{\rm C}$ are the Lindblad and corotation torques, respectively. The Lindblad and corotation torques strongly dependent on the local radial gradients of surface density, $ \Sigma_{\rm g} \propto r^{-\lambda} $, temperature $T \propto r^{-\beta} $, and entropy $S \propto r^{-\varepsilon} $, with $ \varepsilon = {\beta} + \left(\gamma - 1.0\right)\lambda $ and $\gamma = 1.4  $ is the adiabatic index. When the planet grows massive enough, it carves gap in the disc \citep{Crida2006} and eventually switches to the slower type - II migration \citep{Baruteau}:
  \begin{equation}
      \tau_{\rm II} = \tau_{\rm \upsilon} \times \rm max\left(1, \frac{M_{\rm p}}{4\pi \Sigma_{g} r_{\rm p}^2}\right).
      \label{reducedmig}
  \end{equation}
Here, $\tau_{\rm \upsilon}$ is accretion time scale. We combine the disc model, planet formation and migration to synthesise planet populations by sampling the initial conditions. This is described in detail in the following section.

\subsection{Initial Conditions}

Synthesis of a planet population requires a sampling of important input parameters for planet formation. We study the influence of the initial planet starting position, metallicity and disc lifetime. For the disc lifetime and the starting time of the planetary seeds, we assumed a disc lifetime centred around 3 Myr, with a Gaussian distribution with $\sigma$ = 0.5 Myr, where we fixed the minimal and maximal lifetime of the disc to 2 and 5 Myr, respectively.

The metallicity of the disc follows a Gaussian distribution centred around 0.0678, where $Z_{\rm tot}=Z_{\rm dust}+Z_{\rm peb} = 1.5 \%$ corresponds to the solar value, [Fe/H]=0. Here, $Z_{\rm dust}$ is the metallicity of the dust grains responsible for the disc structure \citep{Bitsch}, which we set to 20\% of the total metallicity $Z_{\rm tot}$. $Z_{\rm peb}$ is the metallicity of pebbles, which corresponds to 80\% of the total metallicity $Z_{\rm tot}$. For the initial planetary seed position, we assumed a linear or logarithmic distribution in semi major axis in the regime of [0.1:50] AU.

\section{Results and Discussion}

\subsection{Growth track}

\begin{figure}
 \centering
 \includegraphics[width=\columnwidth]{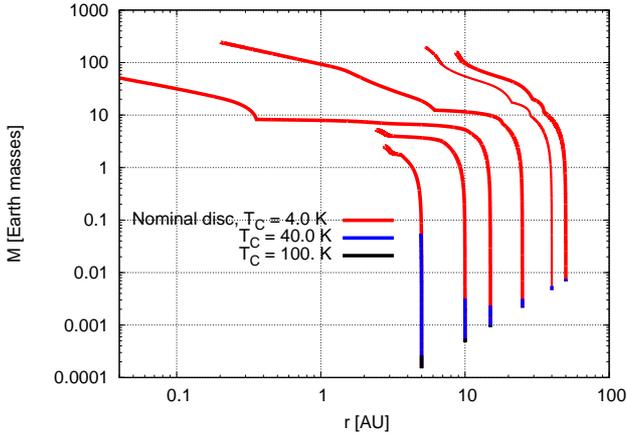}
 \caption{Growth tracks of planets accreting pebbles in a evolving protoplanetary disc submerged in stellar clusters of different heating.
 Different colours represent the planet's growth in different cluster environments. The planets started growing at 2 Myr up to 3 Myr. In the example of red line, the planet reaches 0.04 AU before the disc lifetime and we stopped the simulation. The coloured lines at different r values indicate different types of planet evolution.}
  \label{fig:2}
\end{figure}

To illustrate the growth mechanism, we use a metallicity of $Z_{\rm peb}$ = 1.2$\%$ in pebbles and $Z_{\rm dust}$ = 0.3$\%$ in dust for the disc structure, corresponding to [Fe/H] = 0. All planetary seeds in Fig.~\ref{fig:2} were implanted at an initial time of $t_{\rm 0} = 2 $ Myr and at different initial positions $ r_{\rm 0} $, implying planets will evolve for 1 Myr, because the disc lifetime is 3 Myr. The starting mass of the planets is different  at different positions, because the pebble transition mass is a strong function of $H/r$ and the radial pressure gradient (Equation \ref{eq:transitionmass}). Throughout the study, we used a pebble transition mass obtained for the disc profiles of the isolated field star in order to investigate the different growth mechanisms in detail for the simulations with different cluster temperature. As expected, planetary seeds implanted in clusters with different background temperatures show different growth behaviour, where growth in the outer parts of the disc is significantly reduced in the discs embedded in cluster environments of higher temperature, due to the larger pebble scale height allowing only 3D accretion and the loss of volatile pebbles (CO and CO$_2$).

Planetary seeds do not grow significantly in disc embedded in clusters with $T>40$ K. The slow growth rates are a result of the large pebble scale height, allowing only 3D pebble accretion in combination with the loss of volatile pebbles that evaporate for these high temperatures (CO and CO$_2$). Additionally, the growth time is only 1 Myr. In section~\ref{subsec:synthesis} we vary the lifetime of the protoplanetary disc and the time $t_0$ at which the planetary seeds are embedded in the protoplanetary disc, which results in further growth. In contrast, the planetary seeds embedded in the disc of an isolated star grow rapidly at all orbital distances, even out to $50$ AU, because of the smaller pebble scale height and the larger pebble flux (all volatile species are available). Note here, that the planet growth rates are higher compared to \cite{Bitsch2015}, because the amount of pebbles $Z_{\rm peb}$ is slightly larger.

In the inner disc ($r<10$ AU), the planets forming in protoplanetary discs heated by cluster environments still do not grow. This is a result of the increased disc scale height, the low pebble flux towards the end of the disc's lifetime and the relatively short growth time-scale of 1 Myr. Only in the nominal case, planets can grow. These planets form cores of a few Earth masses that are then parked in the region of outward migration at 2-3 AU. They have reached pebble isolation mass and then slowly contract a gaseous envelope. However, the gas contraction time-scale is longer than the here used growth time of 1 Myr.

%In the inner disc ($r<10$ AU), however, planets grow to only super Earth sizes.  The planetary cores grow too big to be contained in the region of outward migration and drift inwards at masses larger than what the region of outward migration can contain before the planet contracts its envelope and goes into runway gas accretion. These planets all reach the inner edge within the lifetime of the disc, with two exception for the isolated field star, where planet actually stays small enough to be contained in the region of outward migration until disc dissipation. The difference here to \cite{Bitsch2015} is the lower amount of dust grains in the protoplanetary disc ($Z_{\rm dust} = 0.15\%$), which greatly reduces the region of outward migration \citep{Bitsch}.

\subsection{Planet population synthesis}
\label{subsec:synthesis}

Fig.~\ref{fig:2} shows the growth tracks of planetary seeds implanted at different positions in the disc, fixed in initial starting time, disc lifetime and metallicity. Extending Fig.~\ref{fig:2} in the dimensions of initial starting position, disc lifetime and starting time of the planetary seed as well as metallicity, we synthesize a population of 1350 planets by randomly sampling these quantities. We use here an initial starting time of planetary seeds in the interval [0:1] Myr and disc lifetimes of [2:5] Myr. We use RV (Radial Velocity) data retrieved from the exoplanet.org data base in order to compare our results with observations first. We only rely on RV data, because our simulations calculate the mass of the planet, not the planetary radius for which a sophisticated chemical evolution model of the planet's interior would be needed. In all our population synthesis plots, planets at the top left in the plot are gas giants observable by  the RV detection method as defined in \cite{J2010}.

\begin{figure}
        \centering

%        \subfigure{
         \includegraphics[width=\columnwidth]{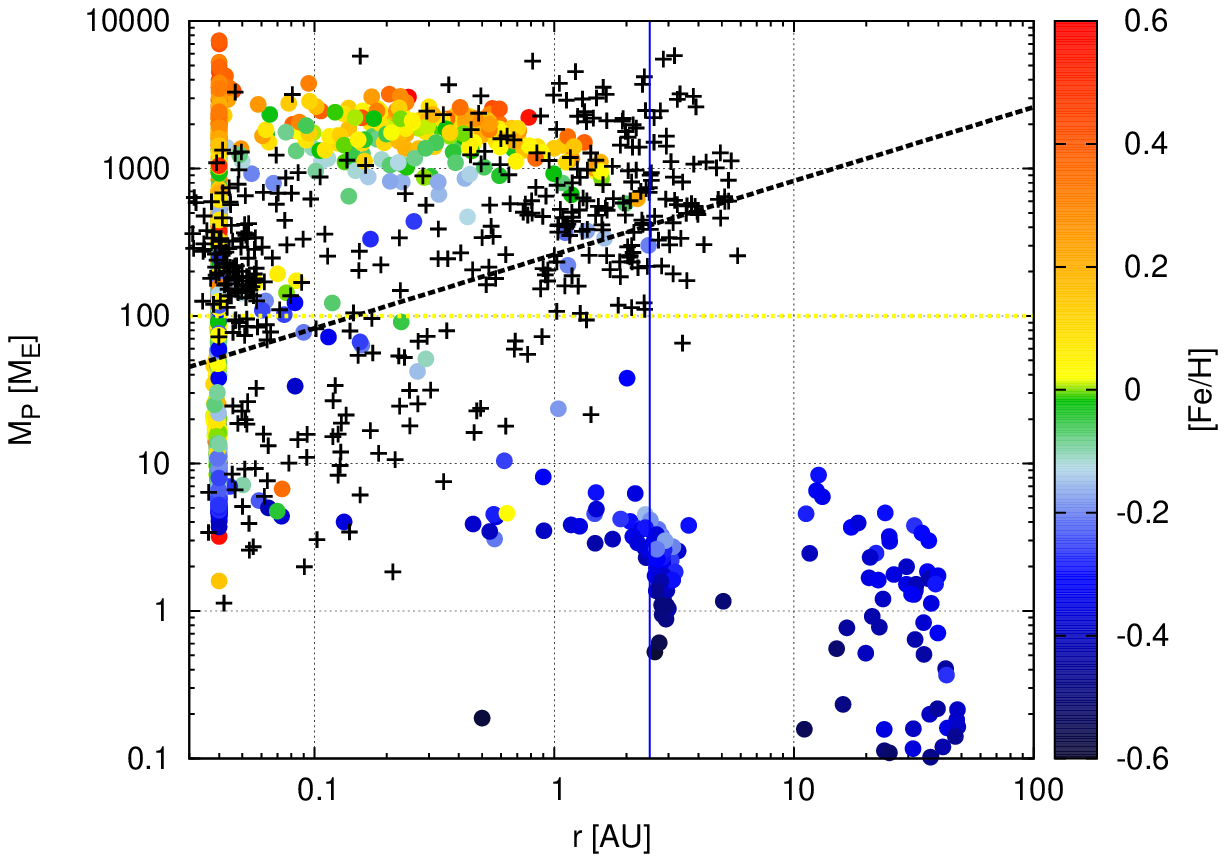}
%         \label{fig:subfigure1}} \hfill
 %        \subfigure{
         \includegraphics[width=\columnwidth]{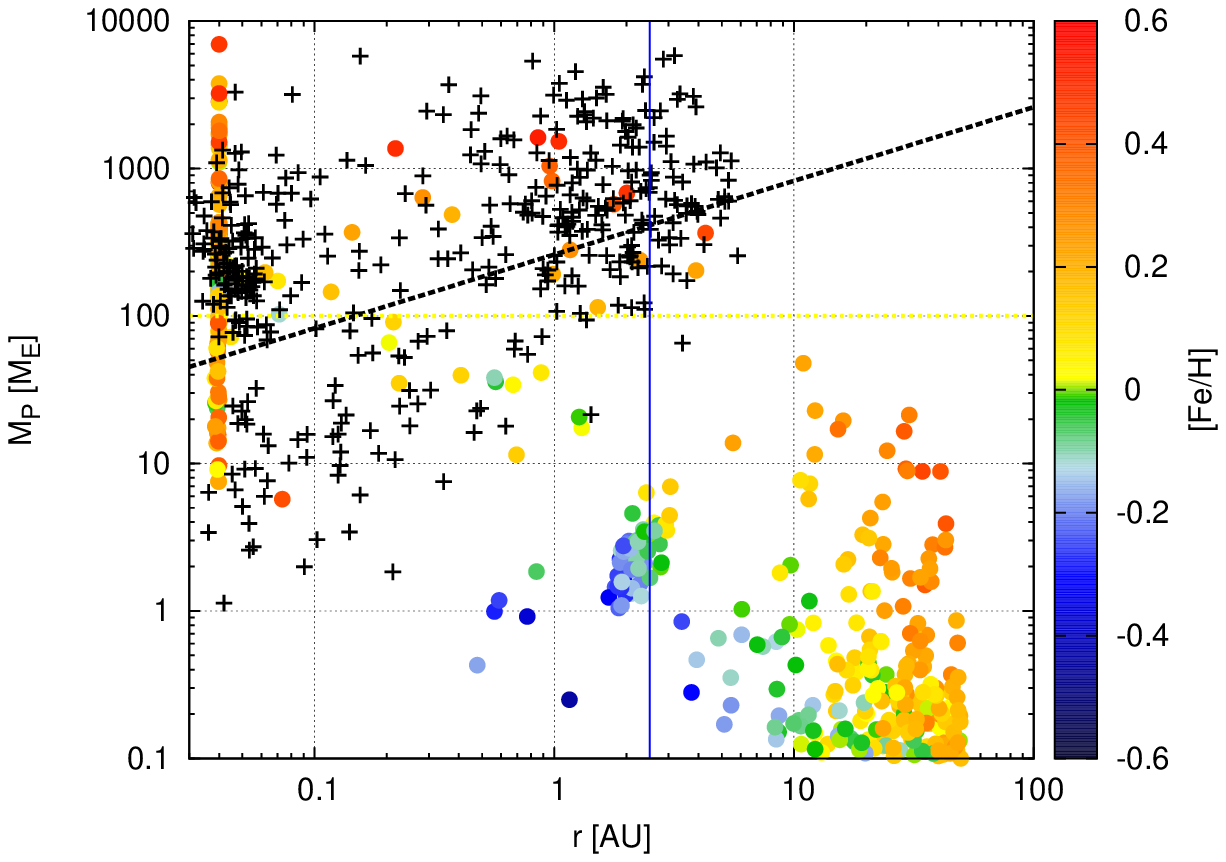}
 %        \label{fig:subfigure2}} \hfill
 %        \subfigure{
         \includegraphics[width=\columnwidth]{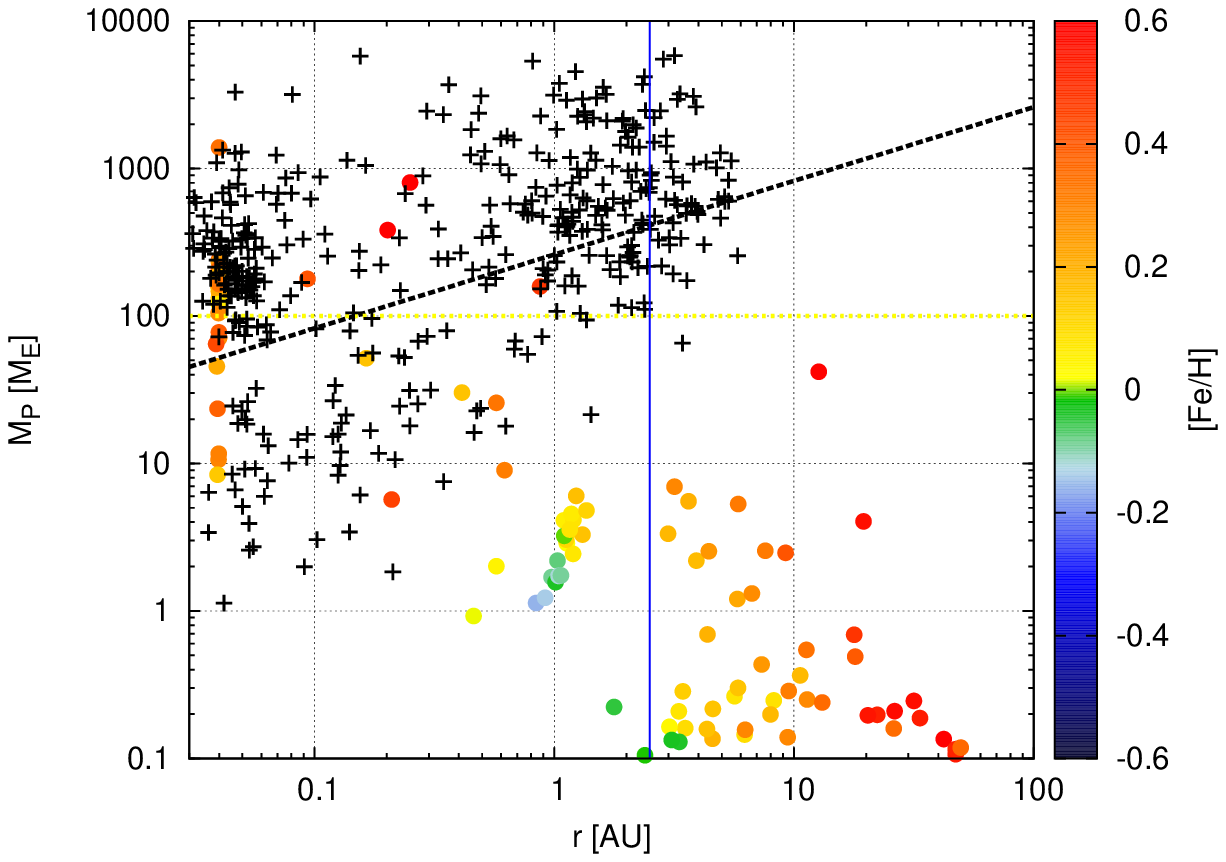}
   %      \label{fig:subfigure3}} \hfill
        
         \caption{Planet population synthesis for cluster of 4 K (top plot), 40 K (middle plot) and 100 K (bottom plot), respectively with initial starting time between 0-1 Myr and with the linear distribution in starting positions. Different colours features different metallicity values. The disc life time spans [2-5] Myrs. The black crosses represent RV data retrieved from the exoplanet.org data base. The diagonal black line marks the RV detectability with $K=20$ m/s. The vertical blue line marks 2.5 AU, which is the radial cut-off for our giant planet frequency calculations and the yellow horizontal line marks 100 Earth masses, which we use as a division between giant planets and no giant planets. Planets inside the top left quadrant represent giant planets that were observed until completeness by the RV survey of \citep{J2010}.}
         \label{fig:3}
 \end{figure}

Fig.~\ref{fig:3} shows the population of the planets synthesized in an isolated stellar field ($ T_{\rm c} = 4 $ K), 40 K stellar cluster and 100 K stellar cluster. Stellar clusters heat up the outer parts of the protoplanetary disc, which sets a higher aspect ratio and consequently a higher pebble scale height, pebble isolation mass and lower pebble surface density. Additionally, some volatile species are not present in the form of pebbles any more, because the temperature is higher than their evaporation temperature (CO and CO$_2$). Forming planets in a disc with high pebble scale height hardly reach the pebble isolation mass, if they grow in the outer parts of the disc, because the planets accrete pebbles in the slower 3D accretion regime due to the increasingly high $ H_{\rm peb}$ in the outer disc. This explains the reduced number of giant planets synthesized in a clustered stellar field, reflected in the middle and bottom plots of Fig.~\ref{fig:3}.

In the top plot of Fig.~\ref{fig:3}, where planets form in a disc around an isolated host star, gas giants form at all metallicities, while clearly larger metallicities are needed for giant planets forming in discs around stars embedded in clusters. For all these simulations, the final masses of cold Jupiters ($r > 1$ AU) can be reproduces quite well, however, the masses of hot Jupiters ($r < 0.1$ AU) are overestimated by up to a factor of 10 compared to the RV data. This is related to the fact of the long planetary evolution time compared to the disc lifetime (in contrast to Fig.~\ref{fig:2}), which allows planets to grow further and become more massive while migrating inwards. The effects of the lifetime of protoplanetary disc on the formation efficiency of giant planets will be discussed in section~\ref{subsec:lifetime}.

Observation of exoplanets have shown that gas giants preferably form at high metallicity, but hardly form at low metallicity \citep{Fischer,J2010}. This is contrary to the synthesized planet population in the isolated star fields (top plot of Fig.~\ref{fig:3}), where we find in total that 32.2\% of our synthesized planets are super-Earths, 8.9\% are Neptunian/Saturnian planets, 49.7\% are gas giants (hot, warm or cold Jupiters) and 9.2\% are low mass planets (see Table \ref{table1}). Even though our simulations have a wide spread of metallicity, it implies that these simulations form too many gas giants compared to observations. We will discuss about the metallicity relationship in detail in section~\ref{subsec:metallicity}. In contrast, the amount of small planets increases for planets forming around stars that are embedded in a stellar cluster (see Table 1). Observations of exoplanets have also revealed that nearly $50\%$ of all stars are orbited by rocky planets within $1$ AU \citep{Mayor2011}, indicating that the isolated field star disc model with a linear starting configuration of planetary seeds is not necessarily likely, as it fails to reproduce the observations.

\begin{figure}
        \centering
    %    \subfigure{
         \includegraphics[width=\columnwidth]{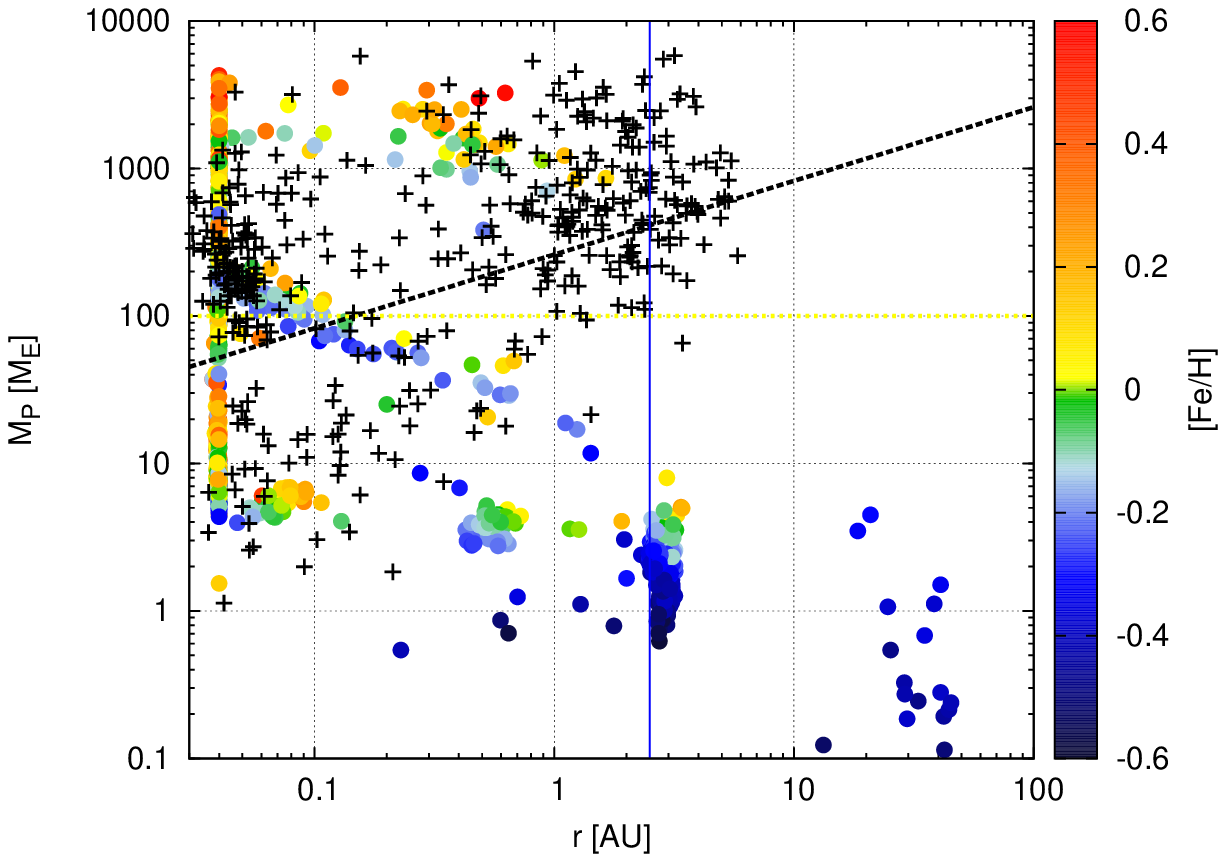}
   %      \label{fig:subfigure1b}} \hfill
    %     \subfigure{
         \includegraphics[width=\columnwidth]{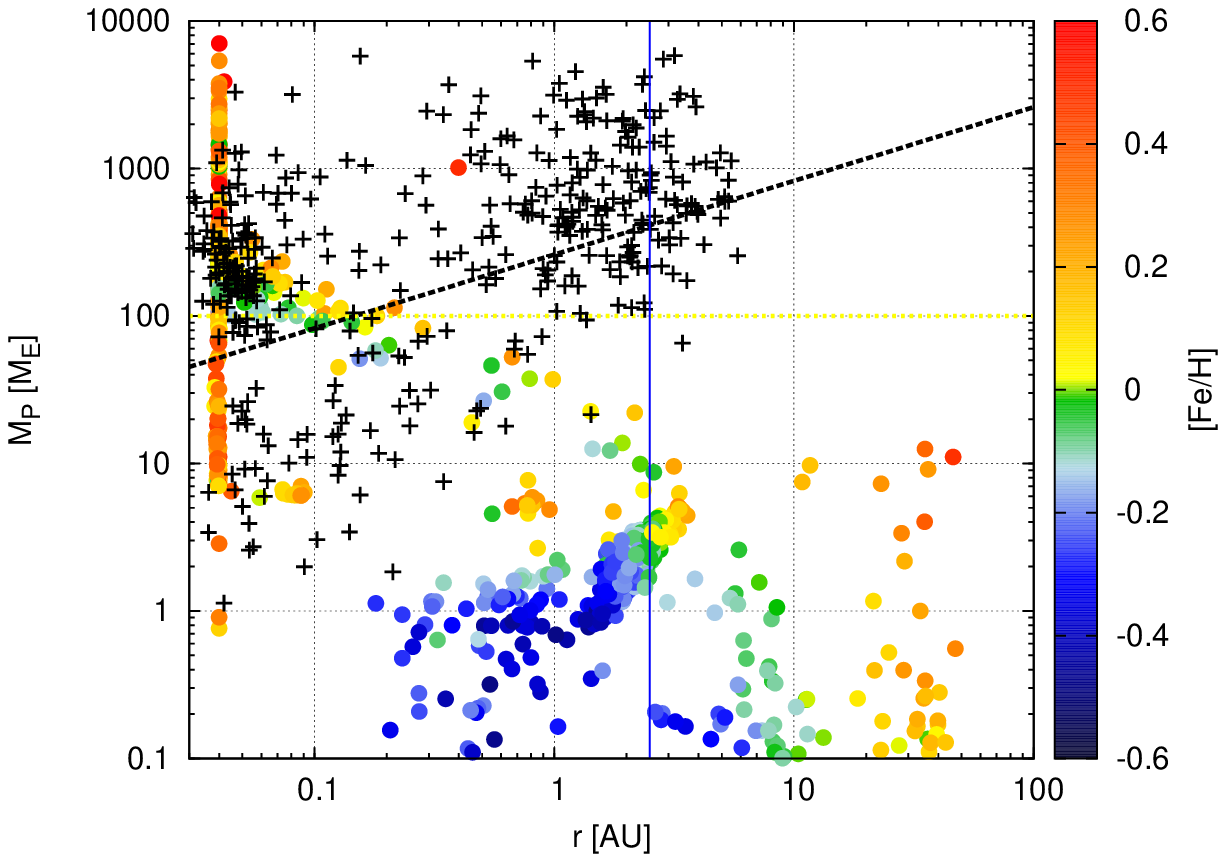}
       %  \label{fig:subfigure2b}} \hfill
         %\subfigure{
         \includegraphics[width=\columnwidth]{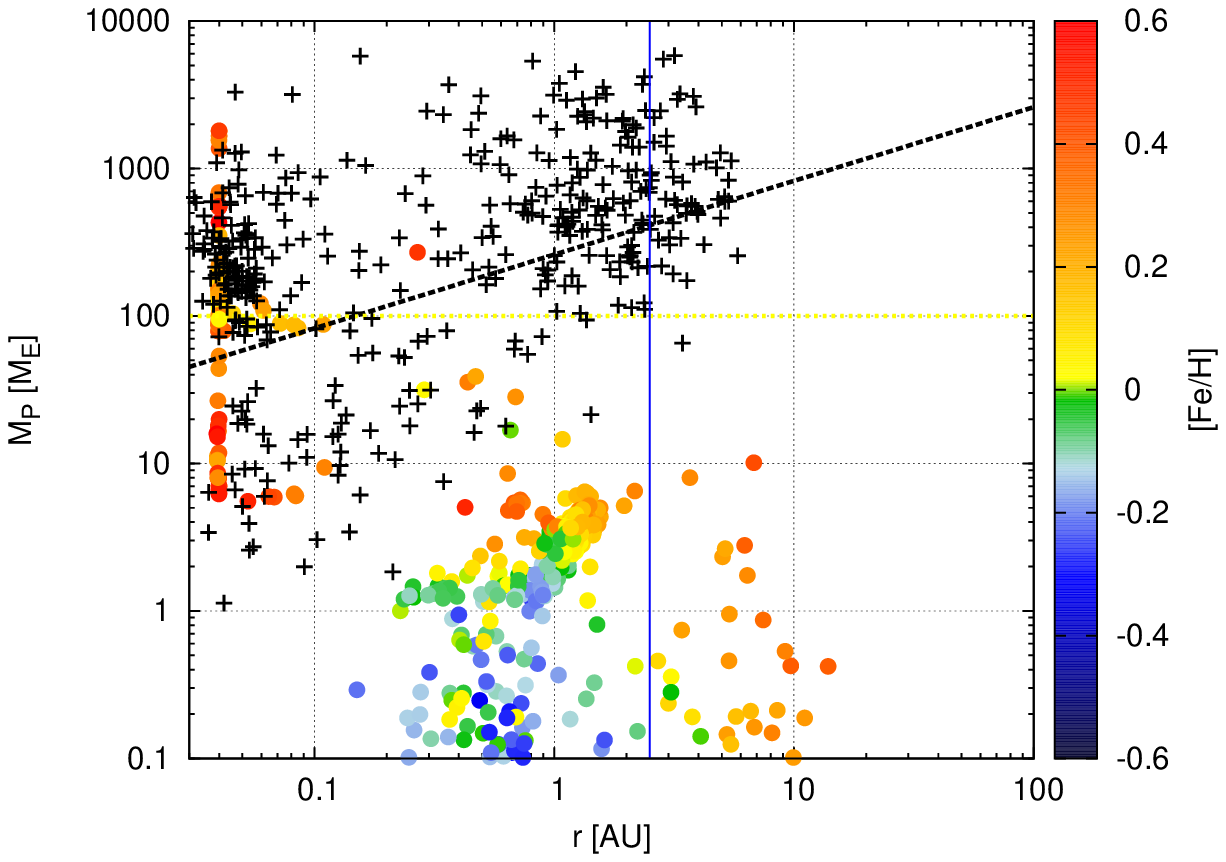}
         %\label{fig:subfigure3b}} \hfill
         \caption{Here different symbols and colours adopt the same meaning as in Fig.~\ref{fig:3}. However, it features a logarithmic distribution in the initial planet starting position contrary to linear distribution in Fig.~\ref{fig:3}.}
         \label{fig:3test}
 \end{figure}

Because the exact starting position of planetary seeds is unknown at the moment, we expand our study to test the influence of a logarithmic radial starting distribution of the planetary seeds and the results are shown in Fig.~\ref{fig:3test}. The maximum mass a planet can grow to is determined by the disc's aspect ratio, because this sets the pebble isolation mass defining the final mass of the planetary core (Equation \ref{eq:Iso}). In the inner parts of the protoplanetary disc, the aspect ratio is smaller (Fig.~\ref{fig:1}) and thus the pebble isolation mass is smaller. As can be seen from Fig.~\ref{fig:2} planetary seeds forming in the inner parts of the disc stay small during their evolution. The logarithmic starting configuration of the planetary seeds therefore results in a much higher super-Earth abundance compared to the linear starting configuration, where more planetary seeds form in the outer parts of the disc. Our findings are that the isolated field star produces a much higher ratio of super Earths, highlighting the importance of the starting position of the planetary seed. Super Earths formation is possible here in contrast to \cite{Bitsch2015}, because of cores forming in the inner disc ($r < 3$ AU). Notably, the pile up of planets at 2-3 AU disappeared for hotter clusters for the population synthesis simulations with logarithmic distribution (see Figure~\ref{fig:3}). This is because hot clusters set higher negative power indices in the surface density gradients which translates into faster inwards migration, destroying the region of outward migration at 2-3 AU, which is present for $T_{\rm c}$ = 4 K.

Additionally, the fraction of hot and cold Jupiters ($r_{\rm p} > 1$ AU) is greatly reduced with increasing stellar cluster background temperatures. This is caused by the reduced core accretion rates for planetary cores forming in the outer disc. However, our simulations still produce a significant fraction of hot Jupiter type planets ($r_{\rm p} < 0.1$ AU), because the seeds for most of the hot Jupiter planets form around 5 AU, where growth is not reduced as much as compared to even larger orbital distances, where some of the hot Jupiters form that migrate all the way to the central star. On top of that, seeds forming in the outer disc have a harder time opening deep gaps and transition to the slower type-II migration, because of the elevated scale height in discs embedded in cluster environments (Fig.~\ref{fig:1}). The few planetary seeds that grow to Jupiter mass in the outer disc are thus lost to the inner disc, increasing the hot Jupiter abundance and decreasing the cold Jupiter abundance.

 {%
 \begin{table}
  \centering
  Linear starting configuration:
  \begin{tabular}{ccccccc}
  \hline \hline
  $T_{\rm c} [K]$ & SE & NP & HJ & WJ & CJ & Others  \\ \hline
4 & 32.2 & 8.9 & 32.4 & 16.1 & 1.2 & 9.2 \\
40 & 5.3 & 3.3 & 10.7 & 0.7 & 0.7 & 79.3 \\
100 & 1.9 & 0.6 & 1.2 & 0.0 & 0.0 & 96.3 \\
  \hline
  \end{tabular}\\
  Logarithmic starting configuration:
  \begin{tabular}{ccccccc}
  \hline \hline
  $T_{\rm c} [K]$ & SE & NP & HJ & WJ & CJ & Others  \\ \hline
  4 & 37.2 & 11.8 & 35.3 & 3.6 & 0.2 & 11.9 \\
  40 & 24.6 & 3.9 & 24.0 & 0.5 & 0.0 & 47.0 \\
  100 & 15.8 & 1.9 & 5.3 & 0.0 & 0.0 & 77.0 \\
  \hline
  \end{tabular}  
  \caption{Proportions of synthesised planets in $\%$ for the linear starting distribution (top) and for the logarithmic starting configuration (bottom). Here SE, NP and others depicts super Earth planet ($2 {\rm M}_{\rm E} < M_{\rm P} < 16 {\rm M}_{\rm E}$ ), Neptunian/Saturnian planet ($ 16 {\rm M}_{\rm E} < M_{\rm P} < 100 {\rm M}_{\rm E}$ ), and Other planet ($M_{\rm P} < 2 {\rm M}_{\rm E}$ ), respectively. HJ, WJ, and CJ stand for hot Jupiters ($r_{\rm p} < 0.1$ AU), warm Jupiters (0.1 AU $< r_{\rm p}<$ 1.0 AU) and cold Jupiters ($r_{\rm p} >$ 1.0 AU), where all Jupiter type planets have $ M_{\rm P} > 100 {\rm M}_{\rm E}$. The population statistics features simulation result for starting time  [0:1] Myrs and disc lifetime [2:5] Myrs, shown in Fig.~\ref{fig:3} and \ref{fig:3test}.
  \label{table1}
  }
 \end{table}
 }%

{%
 \begin{table}
  \centering
  Linear starting configuration:
  \begin{tabular}{ccccccc}
  \hline \hline
  $T_{\rm c} [K]$ & SE & NP & HJ & WJ & CJ & Others  \\ \hline
   &SE&NP&GP&Others \\
4 & 25.4 & 9.8 & 15.2 & 16.2 & 14.8 & 18.6 \\
40 & 5.6 & 1.4 & 2.6 & 0.4 & 0.2 & 89.8 \\
100 & 1.2 & 0.0 & 0.0 & 0.0 & 0.0 & 98.8 \\
  \hline
  \end{tabular}\\
  Logarithmic starting configuration:
  \begin{tabular}{ccccccc}
  \hline \hline
  $T_{\rm c} [K]$ & SE & NP & HJ & WJ & CJ & Others  \\ \hline
  4 & 36.7 & 6.0 & 22.2 & 3.3 & 4.1 & 27.7 \\
  40 & 24.2 & 1.8 & 5.2 & 0.5 & 0.0 & 68.3 \\
  100 & 9.0 & 0.4 & 0.2 & 0.0 & 0.0 & 90.4 \\
  \hline
  \end{tabular}  
  \caption{Here the different symbols carry the same meaning as in Table \ref{table1} but with population statistics featuring simulation result shown in Fig.~\ref{fig:lifetime} and \ref{fig:lifetime1}, where the starting time of the planetary seeds is [1:2] Myr.
  \label{table2}
  }
 \end{table}
 }%

\subsection{Influence of growth lifetime}
\label{subsec:lifetime}
 
The difference between the disc's lifetime and the planetary seed's implantation time sets the growth time of the planet. Planetary seeds implanted early enough in a disc with long lifetime have more time to grow to become massive planets, while cores implanted late in a disc with short lifetimes hardly grow to even an Earth mass in the outer part of the disc, especially if the ambient cluster temperature is high and if the metallicity is low. In principle a longer disc lifetime could compensate for a lower planetary accretion rate (e.g. due to a lower metallicity), but this also implies that the planet has more time to migrate and this might increase the probability of migrating all the way to the central star. In this section, the effect of the growth time on planet formation is explored. 

\begin{figure}
 \centering
 %\subfigure{
         \includegraphics[width=\columnwidth]{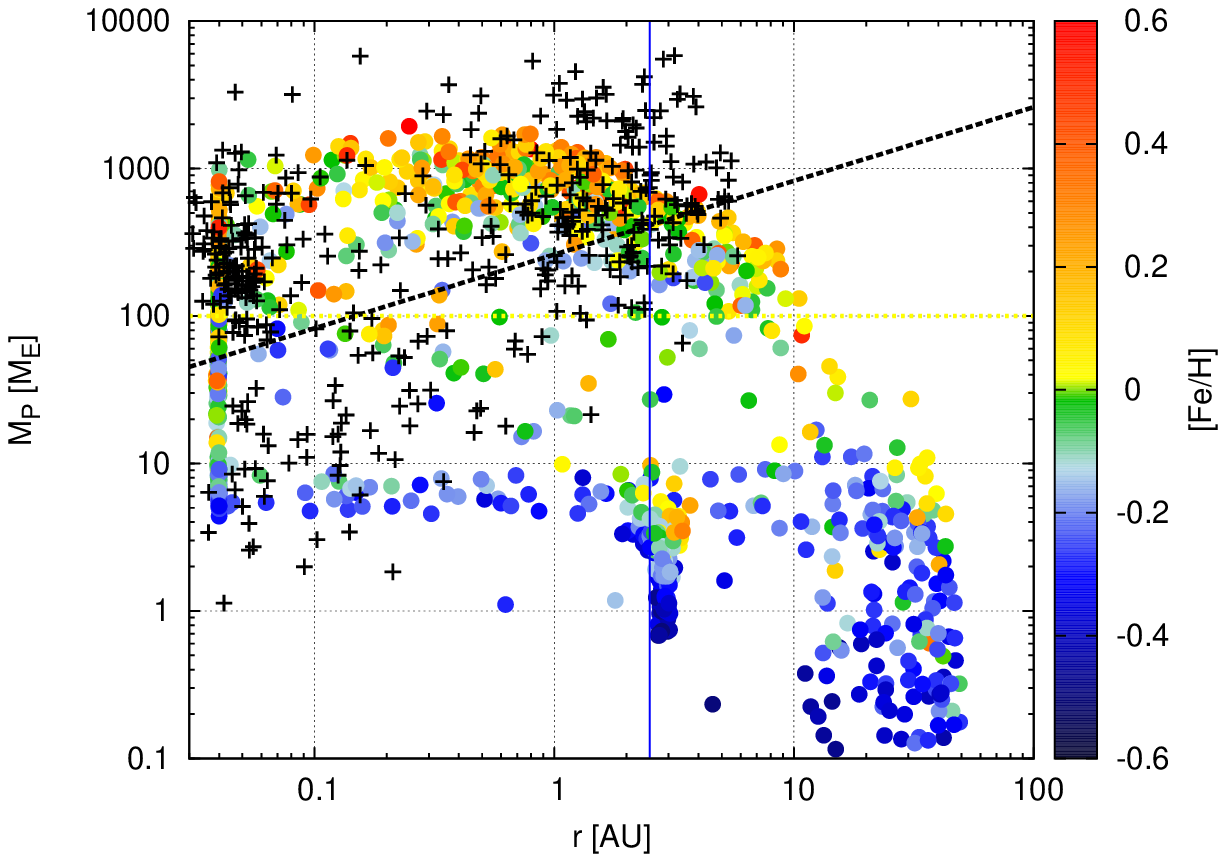}
        % \label{fig:subfigure1c}} \hfill
         %\subfigure{
         \includegraphics[width=\columnwidth]{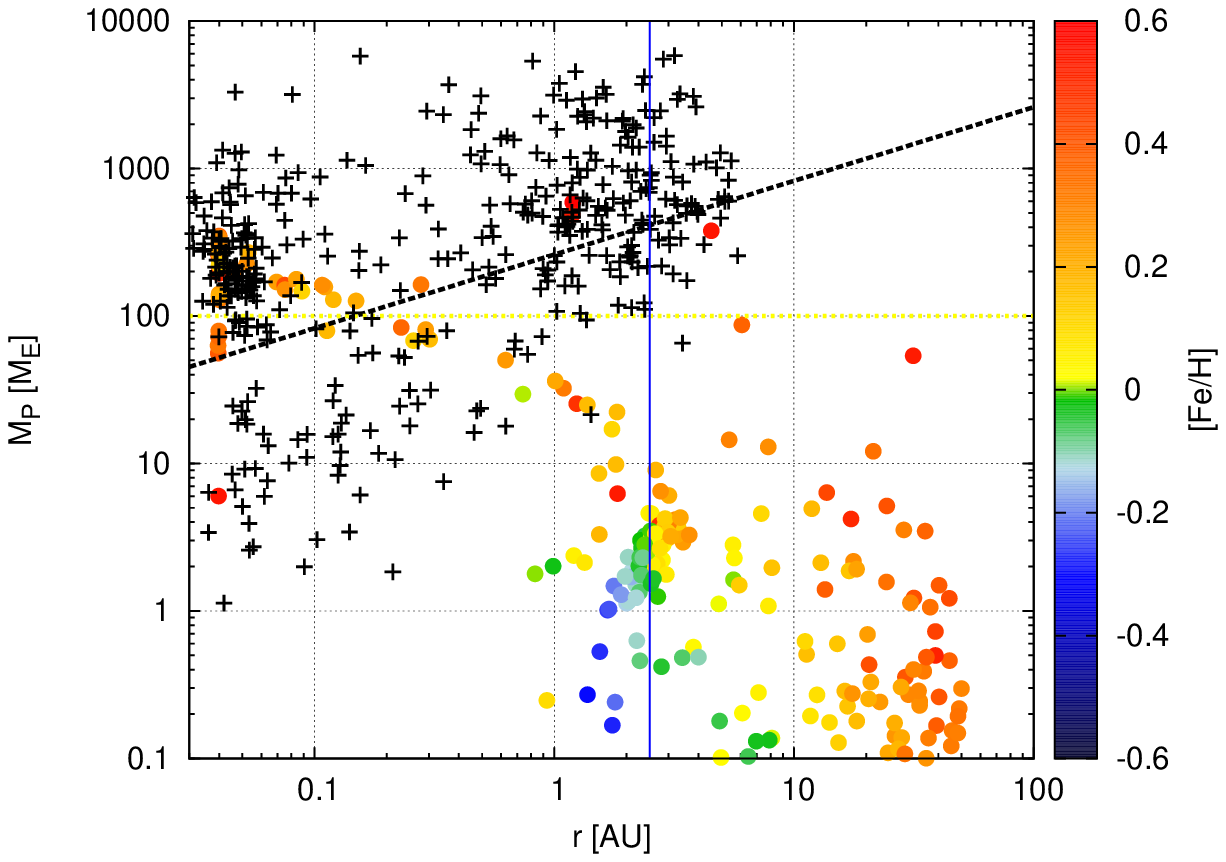}
         %\label{fig:subfigure2c}} \hfill
         
        % \subfigure{
         \includegraphics[width=\columnwidth]{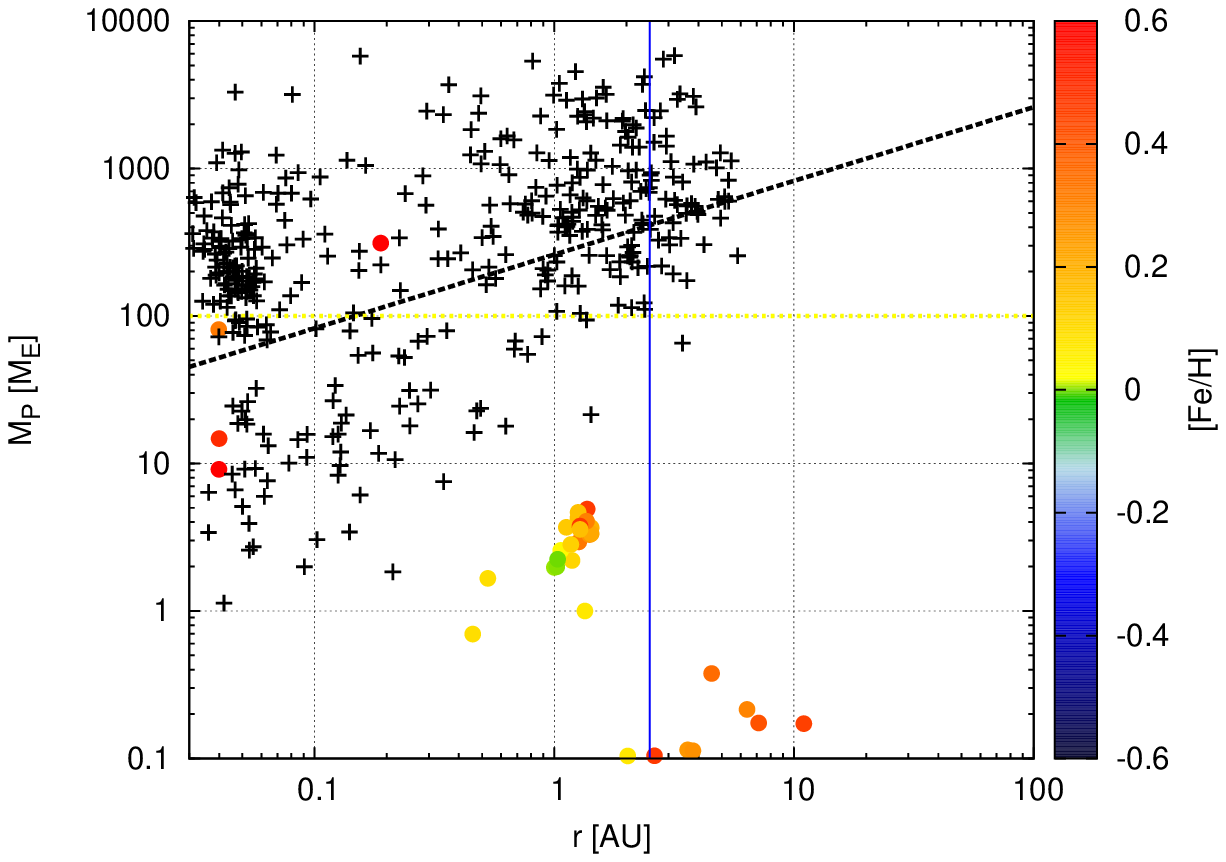}
         %\label{fig:subfigure3c}} \hfill
 \caption{Planet populations synthesized  with later starting times of the planetary seeds [1-2] Myr and linear starting positions. Clearly, the planetary masses for the cold and hot gas giants is lower compared to Fig.~\ref{fig:3}, indicating that the growth time is crucial for the final planetary mass.
   \label{fig:lifetime}
   }
\end{figure}

\begin{figure}
 \centering
%\subfigure{
         \includegraphics[width=\columnwidth]{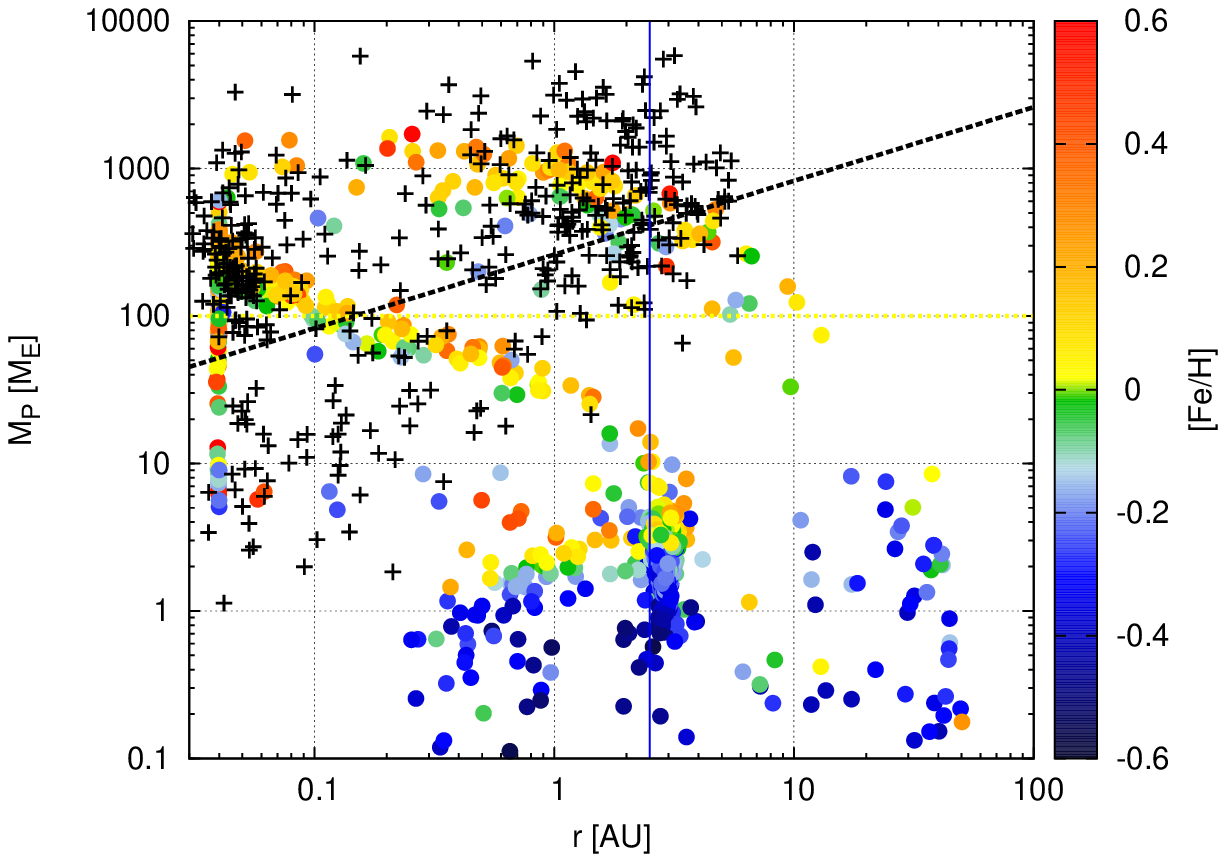}
         %\label{fig:subfigure1d}} \hfill
       % \subfigure{
         \includegraphics[width=\columnwidth]{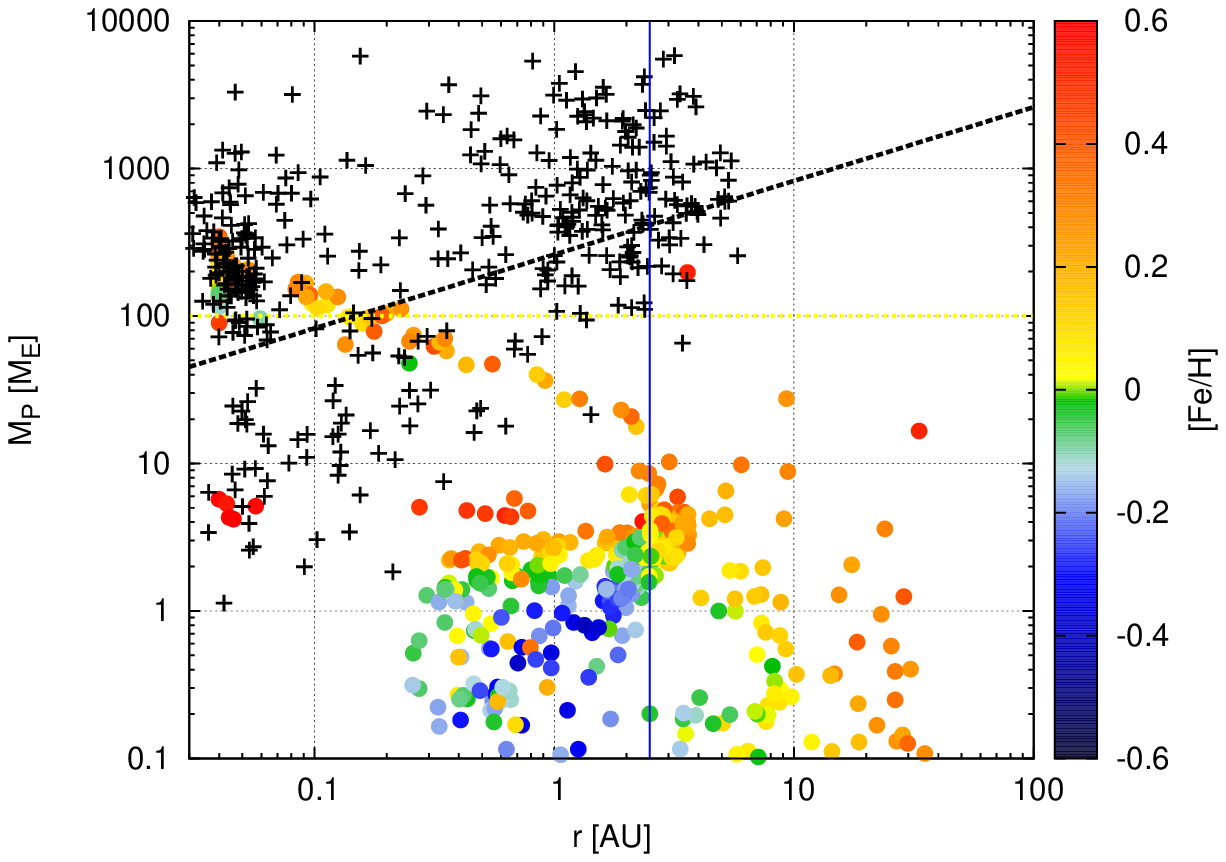}
        % \label{fig:subfigure2d}} \hfill
         
       % \subfigure{
         \includegraphics[width=\columnwidth]{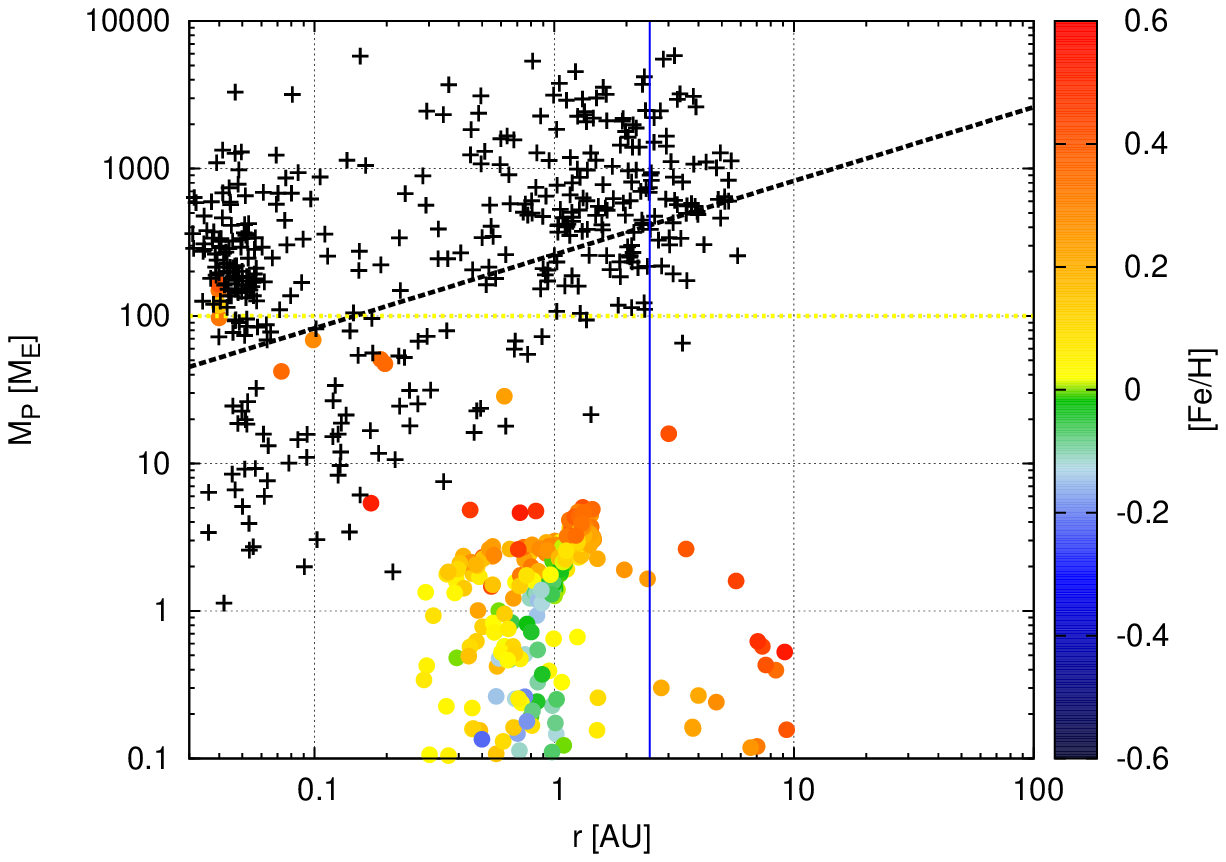}
       % \label{fig:subfigure3d}} \hfill
 \caption{Planet populations synthesized  with later starting times of the planetary seeds [1-2] Myr and logarithmic starting positions. The final planetary masses for the cold and hot gas giants is here also lower compared to Fig.~\ref{fig:3test}.
   \label{fig:lifetime1}
   }
\end{figure}

In Fig.~\ref{fig:lifetime} and Fig.~\ref{fig:lifetime1} we show the influence of the starting time of the planetary seed on the synthesized planet population. Here, planetary seeds are implanted in a disc that is already [1-2] Myr old, so the planetary seeds have less time to grow compared to Fig.~\ref{fig:3}. Clearly the final masses of the cold Jupiter planets is reduced compared to an early starting time of the planetary seeds. Additionally, the final masses of the hot Jupiter planets is greatly reduced compared to the simulations with early starting time of the planetary seeds. This is of crucial importance, because it allows a much better match to observations, where very massive planets are rarely observed. This hints that a later starting time of the planetary seed might be needed to match observations. Additionally, it could hint at too fast gas accretion rates.

Surprisingly, the fraction of gas giants is similar for the early and late starting times for planets forming in discs without external heating from stellar clusters. For early starting times, planets forming in the outer disc can grow to Jupiter sizes and have time to migrate all the way to the central star to become hot Jupiters, while later starting times of the planetary seeds result in less growth and also less migration. However, growth is still efficient enough to let seeds formed in the outer parts of the disc grow to become cold Jupiters at a few AU. However, both, cold and hot Jupiters are counted as gas giants. On the other hand, when planets form in discs exerted to external heating, the giant planet occurrence rate is reduced for later starting times of the planetary seeds. Due to the elevated disc's aspect ratio in the outer disc (caused by the external heating of the disc), the growth time-scale increases. A reduced overall growth time results in less giant planets.

\subsection{Metallicity correlation} 
\label{subsec:metallicity}

\begin{figure}
        \centering

     %   \subfigure{
         \includegraphics[width=\columnwidth]{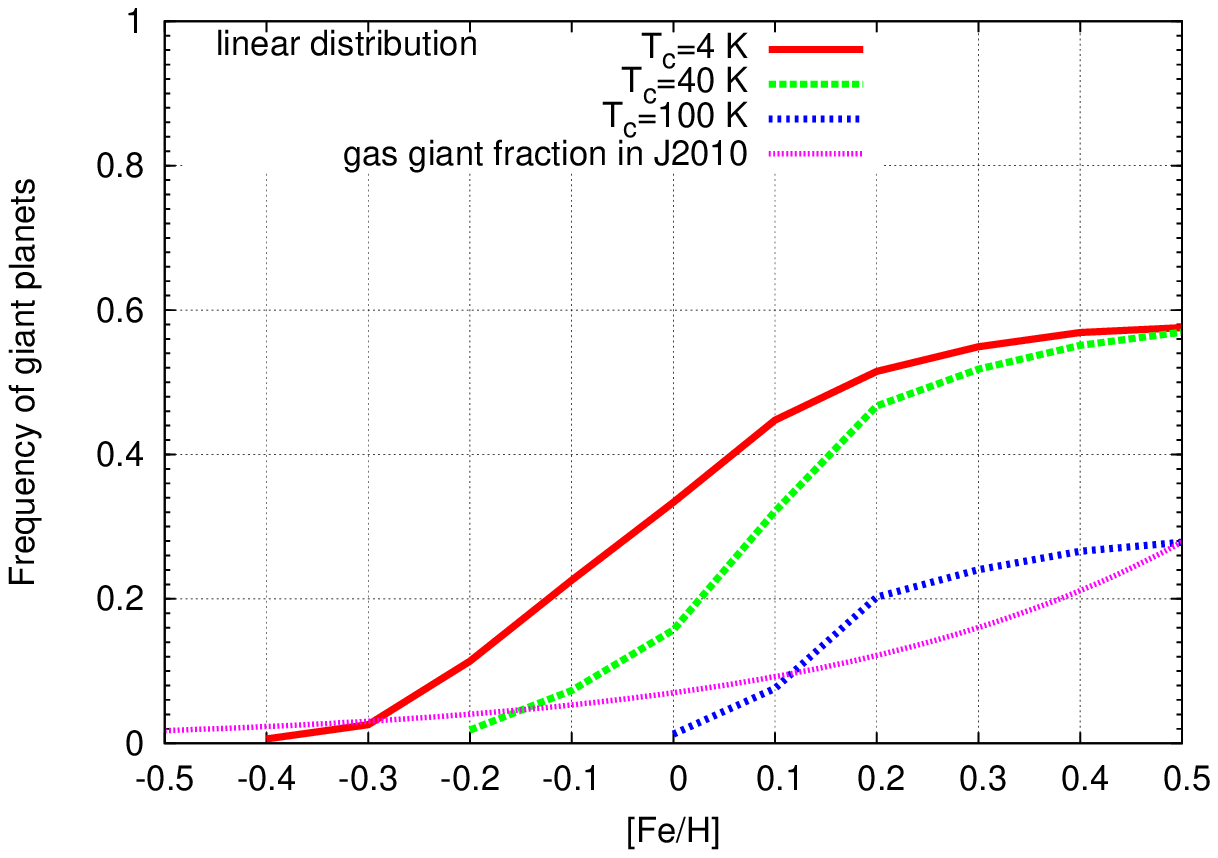}
        % \label{fig:subfigure6a}}\\
        % \subfigure{
         \includegraphics[width=\columnwidth]{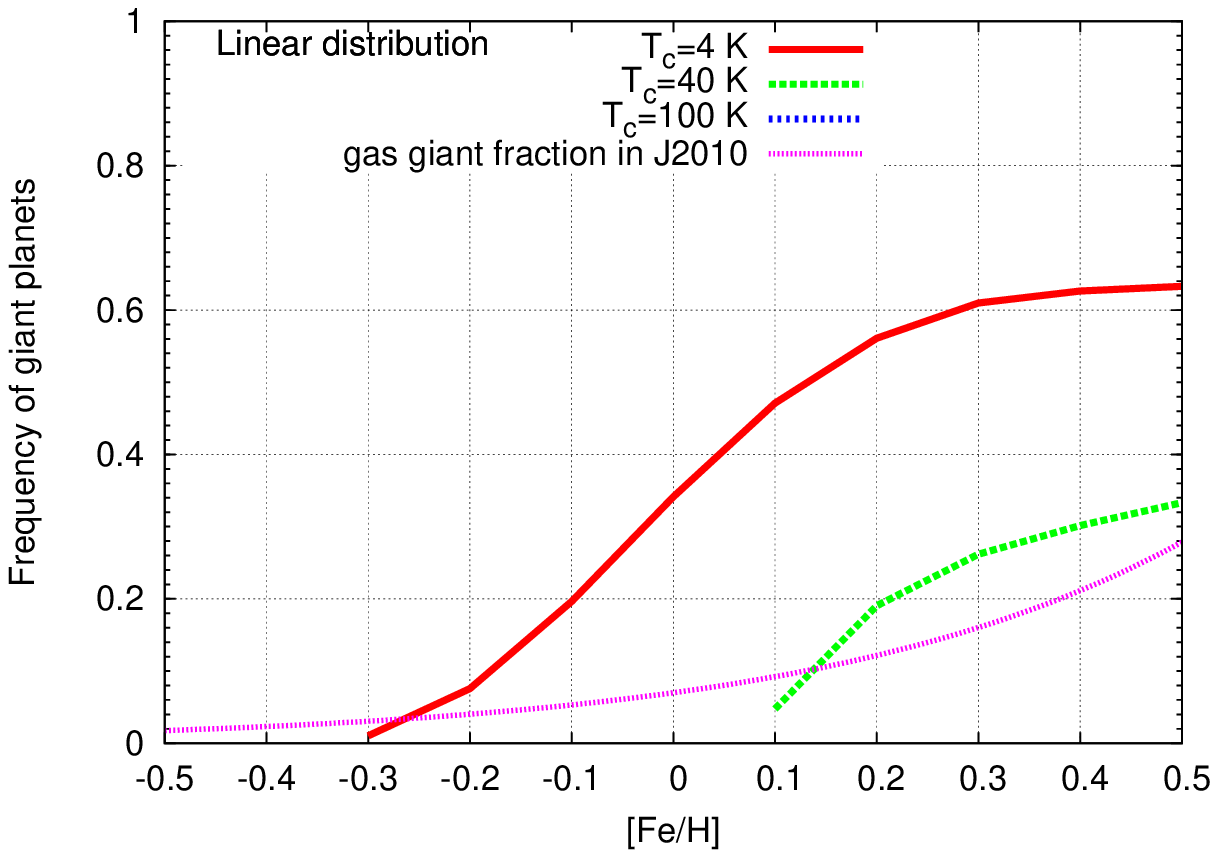}
        % \label{fig:subfigure6b}}
         \caption{Fractions of host stars harbouring gas giants as a function of metallicity. J2010 stands for Johnson et al. (2010), it reflects the curve in Equation \ref{eq:J2010}. For all the clusters, the top plots displays results for  starting time [0:1] Myr and the bottom plot shows result for late starting time [1:2] Myr. All the plots feature the linear distribution in the initial starting position of the planetary seeds.} 
         \label{fig:6}
 \end{figure}

\begin{figure}
        \centering

     %   \subfigure{
         \includegraphics[width=\columnwidth]{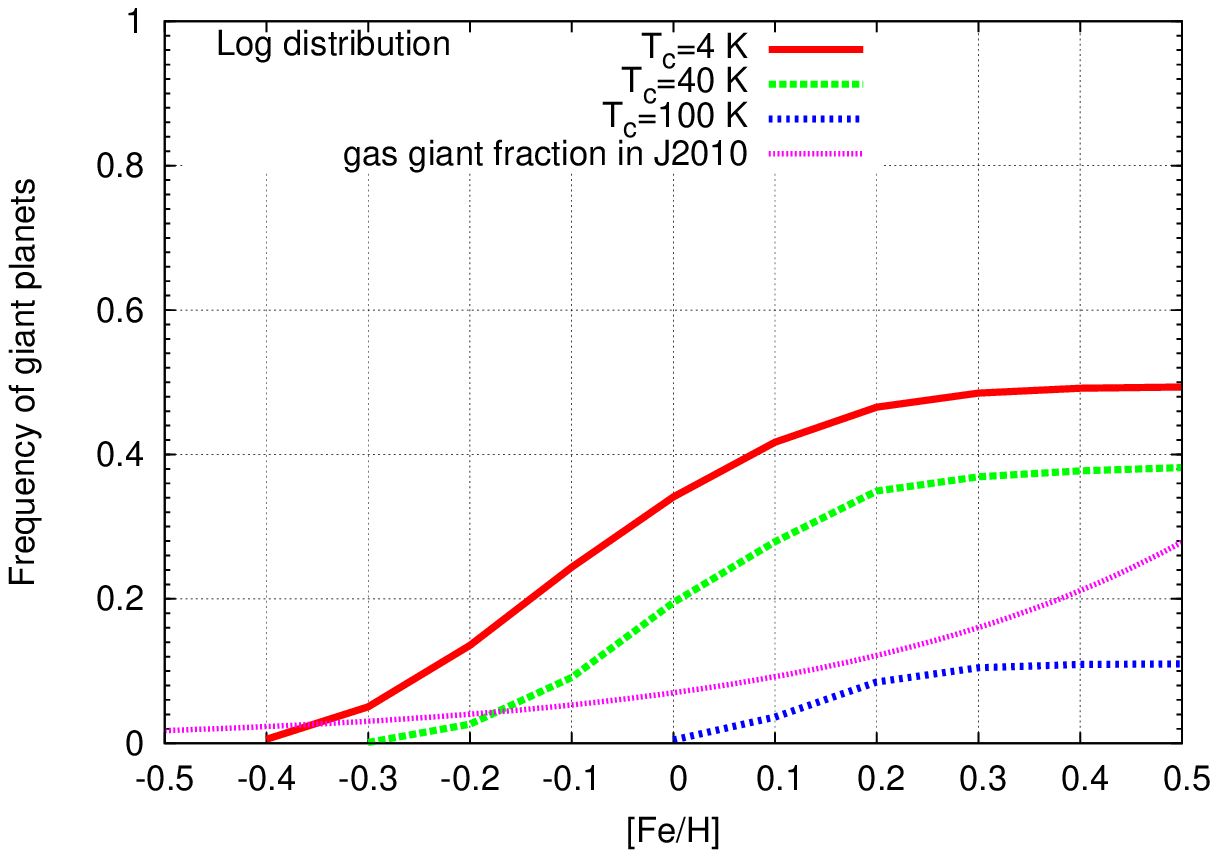}
        % \label{fig:subfigure6a}}\\
        % \subfigure{
         \includegraphics[width=\columnwidth]{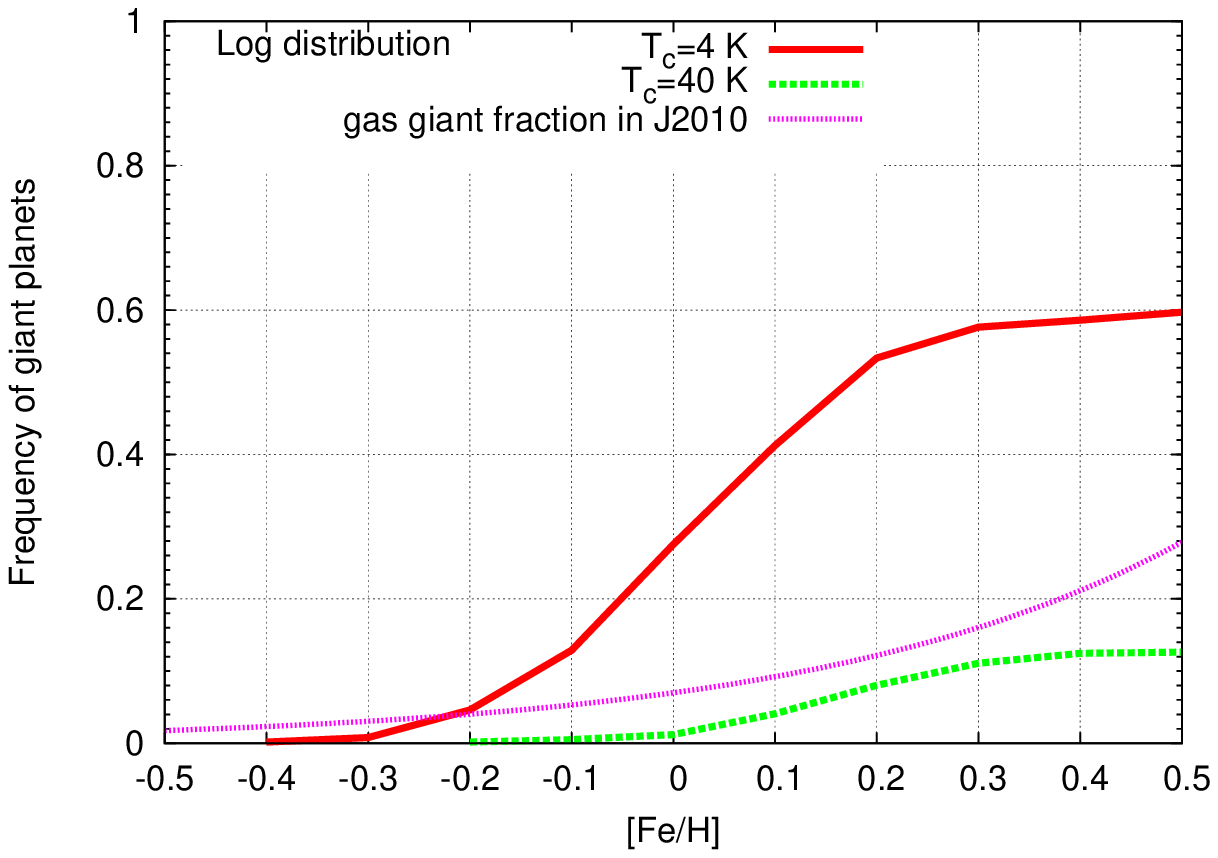}
       %  \label{fig:subfigure7b}}
         \caption{The plots have the same meaning as in Fig.~\ref{fig:6} but for a logarithmic distribution in initial starting position.} 
         \label{fig:7}
 \end{figure}

Observations of giant planets have revealed a strong correlation of giant planets occurrence rate with host star metallicity \citep{Fischer,J2010}. In particular the occurrence rate of giant planets increases with host star metallicity. In the RV observations of \cite{J2010}, the giant planet occurrence rate is given by
\begin{equation}
 f = 0.07 \times (M_\star / M_\odot) \times 10^{1.2[\rm Fe/H]} \ ,
 \label{eq:J2010}
\end{equation}
indicating that $7\%$ of planets orbiting a solar mass star with solar metallicity are giants. Here we only investigate the planet formation process around solar type stars. In the sample of \cite{J2010} a giant planet was defined as a planet with velocity semi-amplitudes $K > 20$ m/s within $r < 2.5$ AU. This corresponds to a planet of $100 {\rm M}_{\rm E}$ at $0.15$ AU. We now consider from our simulations only planets that have reached the inner disc within $2.5$ AU  with Masses $ > 1 {\rm M}_{\rm E}$  and calculate how many of those planets are giant planets ($M_{\rm P} > 100 {\rm M}_{\rm E}$) compared to small planets as a function of metallicity\footnote{We use the $K> 20$ m/s criterion as a limit to define a giant, but with a minimum mass of 100 ${\rm M}_{\rm E}$.}. The results are shown in Fig.~\ref{fig:6} and Fig.~\ref{fig:7} for the linear and logarithmic starting configuration of planetary seeds and for the two different starting times [0-1] Myr and [1-2] Myr, respectively. Additionally we over plot the giant planet occurrence rate found by \cite{J2010} and shown in Equation \ref{eq:J2010}. We have calculated the giant planet fraction in our simulations by taking the ratio between giant planets to small planets within $2.5$ AU.

Clearly, the efficiency of giant planet formation in the pebble accretion scenario increases with increasing metallicity, for both initial starting configurations. High metallicity allows faster accretion rates (Equation \ref{eq:pebbleflux}), which consequently allow cores to reach the pebble isolation mass at an earlier stage, giving more time to contract the gaseous envelope around them and therefore results in an earlier transition to runaway gas accretion, enhancing the formation frequency of giant planets. This results in a very efficient pathway to form giant planets. 

Giant planet formation is very efficient in the pebble accretion scenario for all cluster temperatures, if the seeds form early. However, the formation frequency of hot Jupiters ($r_{\rm p} <$ 0.1 AU compared to cold Jupiters ($r_{\rm p} >$ 1.0 AU) increases significantly with increasing ambient temperatures of stellar clusters. At late formation times, the picture is similar that hot Jupiters are more common in our simulations compared to cold Jupiters at large ambient cluster temperatures, but at the same time, giant planet formation is greatly inhibited for large ambient cluster temperatures as well (Fig.~\ref{fig:3} - Fig.~\ref{fig:lifetime1}).

%It seems that the planet formation frequency is too efficient in the pebble accretion scenario for all cluster temperatures for the early seed implantation time, while a late seed implantation time seems to give a good match to the observations for the high cluster temperature of $T_{\rm c}= 100$ K for the linear starting configuration, however, this does not seem realistic as such high temperatures can hardly be achieved in stellar clusters.

%with $T_{\rm c}= 4$ K and $T_{\rm c}= 40$ K as the high metallicity tail is larger than the observed frequency, while a cluster of $T_{\rm c}= 100$ K seems to follow closely the observed gas frequency curve. For all the cluster environments at low metallicity tail, gas giant formation frequency nearly agrees with the observed gas giant frequency. 

Notably, the starting position of the planetary seeds plays an important role in determining the final mass of the planet, because the pebble isolation mass depends on the position in the disc, where a larger pebble isolation masses can be reached in the outer parts of the protoplanetary disc because of the larger aspect ratio. Planets forming in the outer parts of the disc can thus form cold Jupiters, where their formation is preferred over hot Jupiters at late starting times. Planets forming in the inner disc can only grow to become hot Jupiters because of the rapid inward migration, which prevents planetary seeds to stay at large orbital distances. Therefore the logarithmic starting configuration results in a lower frequency of cold Jupiters compared to the linear starting configuration of planetary seeds (see Fig.~\ref{fig:6} and Fig.~\ref{fig:7}).

%Our simulation extended to 0.04 AU. When a planet reaches this inner edge, the simulation was stopped. Additionally to that, late starting times seems to follow the observed gas giant frequency much better than the early starting time, for all the cluster simulations.\\ 

Our simulations use a one-planet-per-star approach, while most exoplanetary systems host multiple planets. The gravitational interactions between multiple planets can change the formation patterns of planets quite significantly, where an increase of eccentricity reduces the ability to migrate outwards \citep{BitschKley2010} and reduces the accretion rate via pebbles \citep{J2015}. In particular a reduction of the accretion rate might slow down the evolution of planets significantly and thus prevent them to reach the pebble isolation mass and therefore they will not accrete a gaseous envelope and not become gas giants. Additionally, if multiple planets are present, scattering events can take place, reducing the number of planets and influencing the statistics \citep{Sotiriadis2017}. We will discuss about different parameters that influence the results of our model in section~\ref{subsec:caveats}.

%In our simulations we considered one-planet-per-star, we then calculated the giant planet fraction by dividing the number of synthesized giant planets within 2.5 AU by the total number of planets within the same orbital distance. This approach has several odds, since we do not account for scattering and multiplicity which are also important during planet growth. This might affect the outcome of the gas giants both at late stage, where the final system architecture is determined and also at the early stages, where small amounts of eccentricity can significantly reduce pebble accretion \citep{J2015} hindering planetary seeds from reaching pebble isolation mass and become gas giants. Therefore the results of our simulations represent a maximum giant planet fraction per planetary seed per disc.

In summary, we stress the importance of the following points:
\begin{itemize}
 \item The pebble accretion scenario is in agreement with the metallicity correlation for giant planets, meaning that more giant planets are formed at higher metallicity.
 \item The initial position of the planetary seeds determines the faith of giant planet formation in our simulations and if a planet can grow to become a gas giant. In particular, a larger aspect ratio in the outer parts of the disc (due to the ambient cluster temperature) reduces the formation efficiency of giant planets, because of the reduced pebble accretion rates. This results in large differences in the metallicity correlation between the planets forming in discs embedded in clusters with different ambient temperatures. Additionally, giant planets formed in discs with large ambient temperatures due to cluster heating form more hot Jupiters than cold Jupiters.
 \item The initial starting time of the planetary seed influences the growth of the planet itself as a longer evolution time allows  cores to grow to larger planetary masses. It is therefore important when and where a planetary seed forms, which has to be taken into account in future N-body simulations that study the evolution and formation of planetary systems.
\end{itemize}

\subsection{Discussion}
\label{subsec:caveats}
 
Our planet formation model is a one-planet-per-star approach, where we probe the formation of giant planets in discs embedded in clusters with different temperatures. The key result of our model is that an increase of the disc's temperature in the outer parts reduces the formation frequency of giant planets and allows in principle a better match to the metallicity correlation of giant planets. However the cluster temperature also changes the ratio of hot Jupiters compared to cold Jupiters formed in our model. In the following we discuss about different aspects of our simulations.
\begin{itemize}
 \item In reality planetary systems consist mostly of multiple planets and not single planets. In our model we have not taken the effects of multiplicity on the growth of planets into account. Multiplicity is very important, because dynamical interactions of planetary seeds can increase their eccentricity and inclinations, so that the accretion rate as well as the migration rate is greatly reduced \citep{Levison,BitschKley2010}. Additionally many giant planets have a non zero eccentricity, meaning that scattering with other objects has taken place \citep{Sotiriadis2017}.
 \item Stellar clusters do not only influence protoplanetary discs with ambient background temperatures as in our model, but also gravitationally. Fly-bys by other stars in the cluster can destroy protoplanetary discs and significantly change the orbits of planets around the central star. These effects could greatly reduce the giant planet occurrence rates. However, our model treats stellar clusters with densities low enough that encounters are not likely \citep{Rosotti2014}. Additionally, the proximity of protoplanetary discs close to O-stars can drive significant external photoevaporation of discs, changing their structure and evolution \citep{Henney02}. In particular a shorter evolution time results in a reduced giant planet formation rate in cluster environments.
 \item When pebbles drift across ice lines in protoplanetary discs, they lose their volatile component. We have taken this effect into account by reducing the pebble flux whenever an ice line is crossed. However, we did not change the size of the pebbles when they crossed the different ice lines. In \cite{Morbidelli} this effect was used to explain the great dichotomy between the giant planets and the terrestrial planets, where (as one of the aspects) the larger pebble size at $r>r_{\rm ice}$ allows the outer planetary seeds to grow faster and form the core of gas giants of the solar system. However, it is yet unclear how much the pebble size should reduce when a specific ice line is crossed. A reduction of the pebble size would thus slow down the accretion rates, especially in discs around stars embedded in clusters where the ambient temperature is higher than the evaporation temperatures of some species.
 \item Our disc simulations feature only one given $\alpha$ value for the viscosity. A larger viscosity would result in a larger pebble scale height reducing the accretion rate onto the planets and thus lowering the giant planet formation efficiency. A lower viscosity has the opposite effect and reduces the pebble scale height and would allow for faster pebble accretion, presumably enhancing the giant planet formation efficiency. Furthermore, the viscosity determines the migration rates, where a higher $\alpha$-value allows for faster type-II migration and a lower $\alpha$ for slower type-II migration.
\end{itemize}

Our simulations clearly show that cold gas giants form in the outer parts of the protoplanetary disc, while super-Earth type planets and hot Jupiters form in the inner part of the protoplanetary disc. This has important implications also for the chemical composition of planets, which can be traced in planet formation simulations as well \citep{Madhusudhan2017} and can be constraint by observations. In particular for gas giants, \cite{Madhusudhan2017} found that there is a distinct difference in the C/O ratio of the planetary atmosphere depending on the formation location if the planet itself. Future additional observations of exoplanetary atmospheres, also of hot super-Earths, can therefore help to constrain planet formation models and can shine light on the question if the seeds of giant planets form only in the outer parts of the disc, as in our model.

\section{Conclusion}

In this study we combined pebble accretion \citep{Ormel2010,Lambrechts2012,Lambrechts2014,Lamb2014,Morbidelli,Bitsch2015}, gas accretion \citep{Pisso,Machida} and planet migration \citep{Crida2006,CridaMorbidelli2007,Paardekooper2011,Baruteau} in a disc model, extrapolated from the disc structure of \cite{Bitsch}, taking into account the background heating by the stellar cluster. We used this model to synthesize a planet population that allows us to compare our results to observations.

Our planetary seeds start at the pebble transition mass (between $10^{-4}$ and $10^{-2} {\rm M}_{\rm E}$) and grow by subsequent accretion of pebbles. As soon as the planet reached the pebble isolation mass, the flux of pebbles onto the planet is stopped and the gaseous envelope can contract, where the contraction rate is dependent on the mass of the planetary core. In this scenario, larger planetary cores allow for a more rapid envelope contraction, giving more time to the growing planet to reach the runaway gas accretion phase and become a gas dominated planet. The final core mass depends here on the initial position of the planetary seed, because the pebble isolation mass is a strong function of the disc's aspect ratio. In our case this means that planets have to reach larger core masses in the outer regions of the protoplanetary disc due to the flaring nature of the disc's structure.

The entire accretion phase occurs concurrently with planet migration. During the initial stage of planet formation, the planet grows by accreting pebbles while exerting negligible perturbation onto the surrounding disc. At this stage, the planet undergoes type-I migration. Planets in the outer disc undergo inward migration due to the shallow radial gradients in temperature, while planets in the inner disc may migrate outwards. However, the region of outward migration evolves in time as the disc sheds materials to the central star \citep{Bitsch}. As the planets grow in time, they may outgrow the region of outward migration and drift inwards, where their migration is stopped at the inner edge of the disc. Planets that reach the runaway gas accretion phase can become massive enough to open deep gaps in the disc, transitioning to the slower type-II migration.

Within stellar clusters, disc sizes are strongly limited by stellar encounters at stellar densities exceeding $\sim  2 - 3 \times 10^{3}\rm pc^{-2}$ \citep{Rosotti2014}. In our case, we limited our model to just including the ambient heating from the cluster with temperatures $T_{\rm c} \leq 100$ K \citep{Krumholz2006}. According to our background heating estimate, a heating of 100 K is not that easily attainable. However, heating of such magnitudes are seen in \cite{Spaans1995} and \cite{Krumholz2006}. Therefore, we state here that the cluster with ambient temperatur of 100 K is the extreme case. Our disc model does not account for close stellar fly-bys that can disrupt the disc structure and for external photoevaporation from close by stars.

The seeds of cold gas giants have to form in the outer regions of the protoplanetary disc in our simulations, because a larger aspect ratio is needed to form a big enough core that allows runaway gas accretion. For single field stars, this results in a metallicity correlation for giant planets that is much larger than predicted from observations \citep{J2010}. Additionally, the fraction of small planets is lower than the predictions of observations, where roughly $50\%$ of stars should have planets of super-Earths size \citep{Fressin2013}. However, in cluster environments the temperature in the outer parts of the disc can be higher than the evaporation temperatures of some volatiles (e.g. CO), which reduces the amount of material available to form the pebbles needed to form giant planets at those distances. This reduces the accretion rates and results in a smaller giant planet occurrence rate per star for stars embedded in stellar clusters. However, on the other hand this means that as the fraction of giant planets decreases, the fraction of small planets increases, which is more consistent with observations.

Our simulations reproduce the metallicity correlation for giant planets, meaning that more giant planets are formed around stars with larger metallicity. However, in the standard pebble accretion model \citep{Lambrechts2014,Bitsch2015}, the formation frequency of giant planets is very efficient and might even overpredict the formation efficiency of giant planets. As the pebble accretion rate crucially depends on the disc's aspect ratio, a change of the outer disc structure influences the formation frequency of giant planets. Protoplanetary discs around stars embedded in stellar clusters are subject to the ambient heat from the forming cluster (e.g. \cite{Krumholz2006}), which increases the aspect ratio of the outer disc and thus increases the pebble accretion time-scale and therefore reduces the formation frequency of giant planets. However, not all stars remain during their first few Myr in the stellar cluster environment and might leave the cluster and with it the ambient temperature field before their protoplanetary disc dissipates. Therefore our results reflect a minimum giant planet efficiency for planets formed in a cluster environment with the assumptions that star-star interactions and external photoevaporation play only a minimal role.

Large ambient temperatures allow only for a small fraction of giant planets, unless the metallicity is very large. The formation of super Earths is possible for seeds forming in the inner parts of the disc ($r<3$ AU), where the pebble isolation mass is small, allowing only the formation of cores so small (a few Earth masses) that will not enter into runaway gas accretion during the disc's lifetime. Additionally to that, the disc's metallicity plays an important role, because discs with lower metallicity harbour regions of outward migration in the inner disc that can only contain very low mass planets \citep{Bitsch}, which allows the inward migration of formed super Earths from beyond the ice line \citep{Bitsch2015,Bitsch2016}. In our simulations with logarithmic starting configurations our simulations show a super Earth frequency of about 38\%, in rough agreement with observations \citep{Fressin2013}.

In our model, cold Jupiters ($r_{\rm p} > 1$ AU) form preferably in the outer parts of the disc, due to the large scale inward migration. However, their formation is only efficient at low ambient cluster temperatures, and at late formation times of the planetary seeds, implying that maybe standard gas accretion rates are overestimated. Hot Jupiters ($r_{\rm p} <$ 0.1 AU), on the contrary, can also form in the outer disc, but then they have to form at early times in discs with long lifetimes to allow efficient migration all the way to the central star. Another formation pathway for hot Jupiters is their initial formation in the inner parts of the disc ($r_{\rm initial} \approx 5$ AU), where shorter migration is needed to reach the inner edge of the disc. Additionally, hot Jupiters formed in the inner parts of the disc can also efficiently be formed in discs embedded in clusters with larger ambient temperature. This implies that even with our simple model of cluster environments (ignoring stellar encounters and external photoevaporation), the formation of cold Jupiters is inhibited in cluster environments, while hot Jupiters can form easier. This is at odds with observations, which show a larger frequency of cold Jupiters compared to hot Jupiters, hinting that cold Jupiters do not form in protoplanetary discs embedded in cluster environments. The outer disc structure and the formation location of the planetary seeds is therefore responsible, not only for the giant planet occurrence rate, but also for the formation efficiency of cold and hot Jupiters. Future studies thus have to constrain the evolution of the outer protoplanetary disc in cluster environments and the formation location of the first planetary seeds.

\section*{Acknowledgements}
We thank the International Science Program (ISP) for their financial support. We also thank Michiel Lambrechts for proof-reading the manuscript. We thank an anonymous referee whose comments helped to improve this manuscript.

\bibliographystyle{mnras}
\bibliography{Bibliography}

\label{lastpage}

\end{document}